%% file: double_layer_zeyu_arxiv.tex
\documentclass[aps,prl,preprint,superscriptaddress, nobibnotes]{revtex4-2}

\usepackage{xcolor}
\usepackage{url}

\usepackage{graphicx}
\usepackage{amsmath}
\usepackage{graphicx}
\usepackage{physics}

\renewcommand{\thefigure}{S\arabic{figure}}
\newcommand{\beginsupplement}{%
  \setcounter{figure}{0}\renewcommand{\thefigure}{S\arabic{figure}}%
  }
  
\begin{document}
\input{main_arxiv}

\clearpage
\beginsupplement

\begin{center}
  \textbf{\large Supplementary Materials}
\end{center}
\vspace{6pt}

\input{SM_arxiv}

\clearpage

\end{document}

%% file: main_arxiv.tex
\title{Interlayer Exciton Condensates between Second Landau Level Orbitals in Double Bilayer Graphene}

\author{Zeyu Hao}
\email{zeyuh@berkeley.edu}
\affiliation{Lawrence Berkeley National Laboratory, Berkeley, California 94720, USA}
\affiliation{Department of Electrical Engineering and Computer Sciences, University of California, Berkeley, California 94720, USA}
\affiliation{Department of Physics, Harvard University, Cambridge, Massachusetts 02138, USA}
\author{A. M. Zimmerman}
\affiliation{Department of Physics, Harvard University, Cambridge, Massachusetts 02138, USA}
\author{Kenji Watanabe}
\affiliation{Research Center for Functional Materials, National Institute for Material Science, 1-1 Namiki, Tsukuba 305-0044, Japan}
\author{Takashi Taniguchi}
\affiliation{International Center for Materials Nanoarchitectonics, National Institute for Material Science, 1-1 Namiki, Tsukuba 305-0044, Japan}
\author{Philip Kim}
\email{philipkim@g.harvard.edu}
\affiliation{Department of Physics, Harvard University, Cambridge, Massachusetts 02138, USA}

\begin{abstract}
We present Coulomb-drag measurements on a heterostructure comprising two Bernal-stacked bilayer graphene (BLG) sheets separated by a 2.5 nm hexagonal boron nitride (hBN) spacer in the quantum Hall (QH) regime. Using top and bottom gate control, together with an interlayer bias, we independently tune the two BLG layers into either the lowest ($N=0$) or second ($N=1$) Landau level (LL) orbital and probe their interlayer QH states. When both layers occupy the $N = 0$ orbital, we observe both interlayer exciton condensates (ECs) at integer total filling and interlayer fractional QH states, echoing the results in double monolayer graphene. In contrast to previous studies, however, when both BLG layers occupy the $N=1$ orbital, we also observe quantized drag signals, signifying an interlayer exciton condensate formed between the second LLs. By tuning the layer degree of freedom, we find that this $N=1$ EC state arises only when the $N=1$ wavefunction in each BLG is polarized toward the hBN interface to maximize the interlayer Coulomb interaction. 
\end{abstract}

\maketitle

Electrons moving in two-dimensional (2D) space and under a strong magnetic field fall into discrete Landau levels (LLs), indexed by an orbital quantum number $N = 0, 1, 2...$, with the corresponding spatial wavefunction becoming increasingly nodal as $N$ increases. The ultraflat electronic dispersion creates an ideal environment for Coulomb interactions to dominate and induce various many-body ground states, such as the Wigner crystal~\cite{fertig_properties_1996}, charge density wave~\cite{fogler_stripe_2001}, and most notably, the fractional quantum Hall (FQH) effect~\cite{stormer_fractional_1999}. The orbital character is critical to interaction physics because the Haldane pseudopotential for LL wavefunctions varies with $N$~\cite{haldane_fractional_1983}. In particular, even-denominator FQH states at half-filling have been predominantly observed in the $N = 1$ LL~\cite{GaAs_even_denominator,GaAs_even_denominator2,ZnO_even-denominator_2015,zibrov_even-denominator_2017,li_even-denominator_2017,WSe2_even_denominator}, and are believed to be described by a class of non-Abelian states including the Moore-Read Pfaffian state~\cite{moore_read_state}. Experimentally, access to different LLs effectively serves as a powerful tuning knob for electron-electron interactions, fostering exotic many-body phases.

Quantum Hall (QH) physics can be further enriched by placing two electrically isolated, parallel layers of 2D electron systems close to each other so that electrons in one layer interact solely through Coulomb interactions with those in the other. This double-layer structure enables more complex multi-component states by introducing a new pseudospin—the layer index, and provides unique experimental access to coherent QH states by allowing for probing the two spatially separated pseudospin components, something impossible for a spin or valley based multi-component state. In particular, the multi-component Halperin (111) state ~\cite{halperin_theory_1983} composed of layer pseudospins maps to an interlayer exciton condensate (EC) formed from electrons and holes in the partially filled LLs in the opposite layers~\cite{equivalence_111_exciton_wavefunction,Fertig_wavefunction}. Such interlayer ECs were first observed in GaAs double quantum wells~\cite{GaAs_EC,GaAs_EC2}, and later realized in double layer structures of monolayer or Bernal-stacked bilayer graphene (BLG)~\cite{liu_quantum_2017,li_excitonic_2017,liu_interlayer_2019,li_pairing_2019}. The exceptional flexibility and tunability of these double-layer graphene structures further enabled observation of the BEC-BCS crossover~\cite{liu_crossover_2022}, novel interlayer FQH states~\cite{liu_interlayer_2019,li_pairing_2019} and exciton solids~\cite{zeng_exciton_solid_2023}. 

However, these interlayer correlated states have been observed only when both layers are in the $N = 0$ LL; clear \(N=1\)–\(N=1\) interlayer states remain elusive. Since $N = 1$ can host a different hierarchy of single-layer FQH states, it could likewise support qualitatively new interlayer states in double-layer systems. Furthermore, once an EC between two $N = 1$ LLs is realized, the transition between this interlayer coherent EC and individual-layer even-denominator FQH states could be a highly nontrivial topological phase transition~\cite{Shi_Jain_2008,theory_N_equal1_EC}. Previously, there have been reports of gapped states in 2H-stacked bilayer $\mathrm{WSe_{2}}$~\cite{shi_bilayer_2022} or twisted double bilayer graphene~\cite{Lei_TDBG_2024} that are consistent with interlayer ECs in higher LLs. However, while the absence of a spacer helps achieve strong interlayer coupling, it prevents probing each layer pseudospin separately, making it impossible to conclusively identify the interlayer coherence of EC states. In this Letter, we present the first observation of quantized Coulomb-drag signals between the $N = 1$ LLs in a double bilayer graphene device, providing key signatures of EC formation in higher LLs.

Fig.~1a shows the device image, and Fig.~1b illustrates its layer configuration. Two BLG layers are separated by a 2.5~nm thick hBN spacer and have separate gold contacts. The top and bottom graphite gates control the channel region, while the silicon back gate and additional contact gates ensure good equilibration between the channel region and gold contacts. 

The zero-energy LL of BLG consists of eight quasi-degenerate flavors: spin up and down, valley $K$ and $K'$, and an orbital quantum number, $N = 0$ and 1~\cite{mccann_electronic_2013}. Fig.~1c highlights the valley and orbital flavors. The $N = 0$ orbital wavefunction is similar to the lowest LL in a conventional quadratic-band system like GaAs. By contrast, the $N = 1$ orbital in BLG is an admixture of the conventional lowest and second LL wavefunctions~\cite{mccann_electronic_2013}. Depending on the valley flavor, these wavefunctions are layer-polarized within BLG, illustrated in Fig.~1c. Previous studies have shown that next-order energy anisotropies and Coulomb interactions in BLG induce an energy hierarchy among the eight flavors, leading to their sequential filling as carrier density increases~\cite{hunt_direct_2017,li_even-denominator_2017,zibrov_even-denominator_2017}. A displacement field further modifies the hierarchy by shifting the relative potential within BLG and altering the energetics according to the varying degree of layer polarization. 

In our experiment, by applying DC voltages to the top and bottom graphite gates, $V_{\text{TG}}$ and $V_{\text{BG}}$, as well as between the two BLG layers $V_{\text{int}}$ (with the bottom BLG grounded), we can control the carrier density \(n_{\text{top/bot}}\) and displacement field \(D_{\text{top/bot}}\) in each BLG (see Supplemental Material, hereafter SM~\cite{SM}, Sec.~S1), and selectively tune them into different flavor polarizations. Similar to our previous experiment~\cite{liu_interlayer_2019}, we measure longitudinal/Hall resistances in both drive and drag layers $R_{\text{drive/drag}}^{\text{xx/xy}}$. A nontrivial drag signal often provides a direct indication of an interlayer state. Unless otherwise specified, most of the data presented in the main text were taken with the bottom BLG as the drive layer and the top BLG as the drag layer at temperature $T=250$~mK and magnetic field $B=16$~T.

We first examine the QH phase diagram of the bottom BLG by fixing $V_{\text{TG}}$ to zero and sweeping only $V_{\text{int}}$ and $V_{\text{BG}}$, so that the top BLG effectively serves as a top gate for the bottom BLG. Fig.~1d shows $R_{\text{xx}}^{\text{drive}}$ as a function of $D_{\text{bot}}$, and the LL filling factor, $\nu_{\text{bot}}=n_{\text{bot}}/n_0$, where $n_0 = eB/h$ is the LL degeneracy per unit area. The range of $\nu_{\text{bot}}$ spans from -2 to 4, covering a substantial portion of the zero-energy LL manifold. Generally, we observe vertical features that are consistent with previous studies. The large regions of zero $R_{\text{xx}}^{\text{drive}}$  near integer $\nu_{\text{bot}}$, shown in deep blue, correspond to integer QH states, while the intervening resistive regions indicate partially filled LLs. Two types of partially filled LLs are evident. One has sharp resistance minima indicative of FQH states with a denominator of 3, appearing mainly for $0<\nu_{\text{bot}}<1$ and $2<\nu_{\text{bot}}<3$. This is characteristic of the $N = 0$ LL orbitals in BLG. The other type lacks fractional features and predominantly appears in $1<\nu_{\text{bot}}<2$ and $3<\nu_{\text{bot}}<4$, consistent with occupation of the $N = 1$ LL orbital. 

In addition, there are several resistive horizontal lines consistent with displacement field-induced flavor hierarchy transitions. Prominent examples lie near $D_{\text{bot}}/\epsilon_0\approx0$ across the full \(\nu_{\text{bot}}\) range and near $D_{\text{bot}}/\epsilon_0\approx\pm100~\text{mV/nm}$ around $\nu_{\text{bot}}=1$. Some subtler transition lines become clearer in the numerical derivative of the data (SM, Sec.~S3). In Fig.~1e, we mark these transitions as dashed horizontal lines, together with solid lines indicating the FQH states and boundaries of integer QH states. This phase diagram matches the established BLG flavor transition maps obtained at comparable magnetic fields~\cite{hunt_direct_2017,zibrov_even-denominator_2017}, allowing us to precisely assign the flavor polarization to each LL segment in Fig.~1e (same color coding as Fig.~1c). We also note that the phase boundaries between different LLs exhibit zigzag shapes. This irregularity occurs because as $V_{\text{int}}$ changes, the top BLG intermittently enters its own gapped QH states and momentarily stops to act as a gate.

Setting $V_{\text{int}}=0$, we now tune $\nu_{\text{top}}$ and $\nu_{\text{bot}}$ independently via $V_{\text{TG}}$ and $V_{\text{BG}}$ and study the collective phase diagram of the two BLG layers. Fig.~2a,b show $R_{\text{xx}}^{\text{drive}}$ and the corresponding $R_{xx}^{\text{drag}}$ over $-4\leq\nu_{\text{top}}\leq4$ and $0\leq\nu_{\text{bot}}\leq4$. The vertical stripe features are primarily QH states of the drive layer — the bottom BLG. Based on the aforementioned denominator-3 FQH features, we can already infer that the bottom BLG occupies $N = 0$ LL for $0\leq\nu_{\text{bot}}\leq1$ and $2\leq\nu_{\text{bot}}\leq3$, and $N = 1$ LL for $1\leq\nu_{\text{bot}}\leq2$ and $3\leq\nu_{\text{bot}}\leq4$. This assignment is confirmed by computing the corresponding $D_{\text{bot}}$ (SM, Fig.~S9c) and comparing with the single-layer BLG phase diagram. Similar analysis shows that the top BLG resides in $N = 0$ LL for \(\nu_{\text{top}}\) intervals \([-4,-3]\), \([-2,-1]\), \([0,1]\), \([2,3]\), and in $N = 1$ LL elsewhere. The zigzag shape of the vertical stripes again reflects the quantitative chemical potential change in the top BLG (see discussions in SM, Sec.~S3).

Beyond these intralayer QH features, we also observe interlayer features, highlighted by the eight dashed squares. They arise only when both BLG layers occupy $N = 0$ LL, and appear as diamond-shaped features with a diagonal stripe running along lines of constant integer total filling $\nu_{\text{tot}}=\nu_{\text{top}}+\nu_{\text{bot}}$, where the number of quasi-holes in one layer equals the number of quasi-electrons in the other. For $\nu_{\text{top}}>0$ and $\nu_{\text{bot}}>0$, the diagonal stripes appear as resistance minima approaching zero in both $R_{\text{xx}}^{\text{drag}}$ and $R_{\text{xx}}^{\text{drive}}$, while their \(R_{\text{xy}}^{\text{drag}}\) and \(R_{\text{xy}}^{\text{drive}}\) lock to the quantized value $\frac{1}{\nu_{\text{tot}}}\frac{h}{e^2}$. These signatures are consistent with the ECs in the previous Coulomb-drag studies of double monolayer graphene~\cite{liu_interlayer_2019,li_pairing_2019} and double BLG~\cite{liu_quantum_2017,li_excitonic_2017}. Fig.~2c,d zoom into the region around $(\nu_{\text{bot}}, \nu_{\text{top}}) = (0.5, 2.5)$ at \(B=\) 16~T and 25~T, respectively, and Fig.~2e around \((0.5, 0.5)\), all clearly revealing a wide region of zero $R_{\text{xx}}^{\text{drag}}$ along the diagonal.  An off-diagonal line cut taken along $\nu_{\text{bot}}=\nu_{\text{top}}-2$ (Fig.~2f) shows that exactly at $(\nu_{\text{bot}}, \nu_{\text{top}}) = (0.5, 2.5)$, where the EC occurs, $R_{\text{xx}}^{\text{drag}}=R_{\text{xx}}^{\text{drive}}=0$ and $R_{\text{xy}}^{\text{drag}}=R_{\text{xy}}^{\text{drive}}= \frac{1}{3}\frac{h}{e^2}$. Note that the filling factors in the large-area maps (Fig.~2a,b) are estimated ignoring chemical-potential effects, and thus indicate only approximate state locations. In the zoomed-in maps and line cuts (Fig.~2c–e), where accurate $\nu$ is required, filling factors are calibrated against known fractional QH features (see SM, Sec.~S1 for details).

These diamond-shaped regions also host correlated fractional states of both intra- and interlayer nature. The vertical and horizontal streaks near fractional \(\nu_{\text{top/bot}}\) in Fig.~2c,e correspond to intra-layer FQH states developing independently in each BLG. In addition, there are straight lines of resistance peak/dip features following fractional slopes. They are interlayer FQH states termed semi-quantized states, previously observed in double monolayer graphene~\cite{liu_interlayer_2019} and attributed to electrons binding interlayer vortices via interlayer Coulomb interactions. We defer the detailed analysis of these states to SM, Sec.~S5. The fact that these interlayer FQH states appear in $N = 0$ LL but not in \(N=1\) in double BLG mirrors the behavior of double monolayer graphene ~\cite{liu_interlayer_2019, li_pairing_2019}, suggesting that the interlayer FQH state structure depends more on the LL wavefunction than on the host material itself.

Despite the rich structure at \(V_{\text{int}}=0\), we observe no interlayer states involving N = 1 LLs. This changes when we introduce a finite interlayer bias. Fig.~3a,b show $R_{\text{xx}}^{\text{drive}}$ and $R_{\text{xx}}^{\text{drag}}$ at \(V_{\text{int}}=0.07~\mathrm{V}\). Now the $N = 0$ diamond features, again marked by dashed squares, survive only near four locations: $(\nu_{\text{bot}},\nu_{\text{top}})=$  $(0.5, -1.5\pm2)$ and $(2.5, -1.5\pm2)$. Strikingly, new interlayer states emerge when the bottom BLG lies in $1<\nu_{\text{bot}}<2$ or $3<\nu_{\text{bot}}<4$ and the top BLG in $-3<\nu_{\text{top}}<-2$ or $-1<\nu_{\text{top}}<0$ (dashed circles). Each appears as a narrow diagonal minimum in $R_{\text{xx}}^{\text{drive}}$ along an integer $\nu_{\text{tot}}$, accompanied by a non-zero drag signal, confirming their interlayer nature. Using the single-layer BLG phase diagram, we can identify their locations as both layers occupying $N = 1$ LL. The zoomed-in views overlaying Fig.~3a highlight a clear visual contrast: the $N = 0$ state on the left has a characteristic wide diagonal zero $R_{\text{xx}}^{\text{drive}}$ feature that cuts into the resistive region from one side and is surrounded by other fractional states, whereas the $N = 1$ state on the right appears as a thin zero $R_{\text{xx}}^{\text{drive}}$ line cutting through the resistive region with few nearby features. Raising $V_{\text{int}}$ to $0.1~\mathrm{V}$ reveals still more $N = 1$ interlayer states (SM, Fig.~S11). 

Although visually distinct, the $N = 1$ interlayer states exhibit transport hallmarks of ECs: $R_{\text{xx}}^{\text{drag/drive}}=0$ and \(R_{\text{xy}}^{\text{drag/drive}}=\frac{1}{\nu_{\text{tot}}}\frac{h}{e^2}\). Fig.~3c,e show $R_{\text{xx}}^{\text{drag}}$ and $R_{\text{xy}}^{\text{drag}}$ for the $N = 1$ interlayer state near $(\nu_{\text{bot}},\nu_{\text{top}})$ = $(1.5, 0.5)$ at $V_{\text{int}} = 0.1~\mathrm{V}$. It appears as a thin line of $R_{\text{xx}}^{\text{drag}}$ dip within a narrow resistive strip that runs along the diagonal, and as a peak in $R_{\text{xy}}^{\text{drag}}$ along the same line. A line cut along $\nu_{\text{bot}} = \nu_{\text{top}}-1$ shows that exactly at $(1.5, 0.5)$, $R_{\text{xy}}^{\text{drag}}$ and $R_{\text{xy}}^{\text{drive}}$ approach the quantization value $\frac{1}{2}\frac{h}{e^2}$ (Fig.~3f), while $R_{\text{xx}}^{\text{drag}}$ and $R_{\text{xx}}^{\text{drive}}$ drop to 0 (Fig.~3d), consistent with an interlayer EC. Note that this $N=1$ state is present at $V_{\text{int}}=0.1$~V but absent at $V_{\text{int}}=0.07$~V (Fig.~3b). 

An interesting question is what drives the strong \(V_{\text{int}}\) dependence of the \(N=1\) interlayer EC. The interlayer bias lets us tune \(D_{\text{top}}\) and $\nu_{\text{top}}$ independently while holding $\nu_{\text{bot}}$ constant (see SM, Sec.~S6). Fig.~4a and 4b show 2D maps of $R_{\text{xx}}^{\text{drive}}$ and $R_{\text{xx}}^{\text{drag}}$, respectively, at fixed $\nu_{\text{bot}}=$ 3.5, while Fig.~4c shows $R_{\text{xx}}^{\text{drive}}$ at $\nu_{\text{bot}}=$ 1.5. For these fillings and sweep ranges, the bottom BLG remains mostly in $N=1$ LLs (SM, Sec.~S6). In Fig.~4a and 4c, $R_{\text{xx}}^{\text{drive}}$ exhibits resistive patterns that are generally periodic in \(\nu_{\text{top}}\), reflecting filling of LLs in the top BLG that imprint onto the bottom BLG $R_{\text{xx}}^{\text{drive}}$ via screening effects. Fig.~4d illustrates the partially-filled top BLG LLs as 4 colored stripes bounded by solid lines. More importantly, Fig.~4a and 4c also show four curved boundaries (dashed lines in Fig.~4d) that cut through the repeating LL features near integer $\nu_{\text{top}}$. These curves are consistent with hierarchy transitions in the top BLG, allowing for assignment of the valley and orbital flavor in each LL segment.

Within this parameter space, the N = 1 EC states appear as straight vertical features: drag resistance peaks in the otherwise mostly zero \(R_{\text{xx}}^{\text{drag}}\) (Fig.~4b) that coincide with dips in \(R_{\text{xx}}^{\text{drive}}\) (Fig.~4a,c). Note that an EC should exhibit zero resistance in both \(R_{\text{xx}}^{\text{drag}}\) and \(R_{\text{xx}}^{\text{drive}}\). However, small offsets in \(\nu_{\text{bot}}\) can shift \(R_{\text{xx}}^{\text{drag}}\) away from the narrow resistance dip and produce the peaks in Fig.~4b. In Fig.~4d, we represent the experimentally observed $N = 1$ EC states as thick solid lines. What's immediately clear is that these interlayer EC states occur only when the upper BLG is in the $N = 1$ orbital and \(K'\) valley. Over the same gate range, the lower BLG remains $N = 1$ orbital and \(K\) valley (see SM, Sec.~S6). Being an admixture of the conventional lowest and second LL wavefunctions, the $N = 1$ orbital with \(K'\) (\(K\)) valley in the upper (lower) BLG has its second LL wavefunction component (and larger wavefunction weight) polarized toward the bottom (top) graphene layer within that BLG (upper inset in Fig.~4d). Taken together, these results suggest that the $N = 1$ EC selects a flavor configuration that places the higher-LL wavefunctions in the two BLGs in immediate proximity across the hBN spacer.

Formation of an interlayer EC relies on the interlayer Coulomb interaction and therefore in principle is favored by bringing the active orbitals as close as possible provided there is no significant interlayer tunneling. However, switching layer polarization of a BLG orbital moves the wavefunction by only 0.34~nm, insignificant compared to the 2.5~nm hBN spacer that sets the interlayer Coulomb interaction scale. Geometric proximity alone is therefore unlikely to explain why the \(N=1\) EC selects a specific valley (or layer) flavor combination. A plausible mechanism likely involves the interlayer Haldane pseudopotential, which sensitively depends on the detailed wavefunction distributions. It is possible that the $N = 1$ EC is stabilized only in configurations where the second LL component of the $N=1$ LL in one BLG can interact directly with that in the other, free from screening by the co-existing lowest LL component. 

In conclusion, we have mapped out the interlayer QH phase diagram of double bilayer graphene via Coulomb-drag transport. We observe the formation of interlayer ECs at integer total filling and fractional interlayer QH states when both BLG occupy $N=0$ LLs. Crucially, we report the first observation of interlayer ECs within the $N=1$ LL manifold backed by quantized Coulomb-drag signals. We have also identified the specific flavor combination required for the N=1 EC, revealing the important role of the interlayer Coulomb interaction. Moving forward, reducing the hBN spacer further and pushing to higher magnetic fields may reveal additional interlayer correlated states in the $N=1$ orbital. On the other hand, realizing \(N=1\) interlayer ECs opens a pathway to exploring their potential exotic phase transitions to the intrinsic intralayer even-denominator FQH states in each individual BLG.

\begin{figure*}
    \centering
    \includegraphics{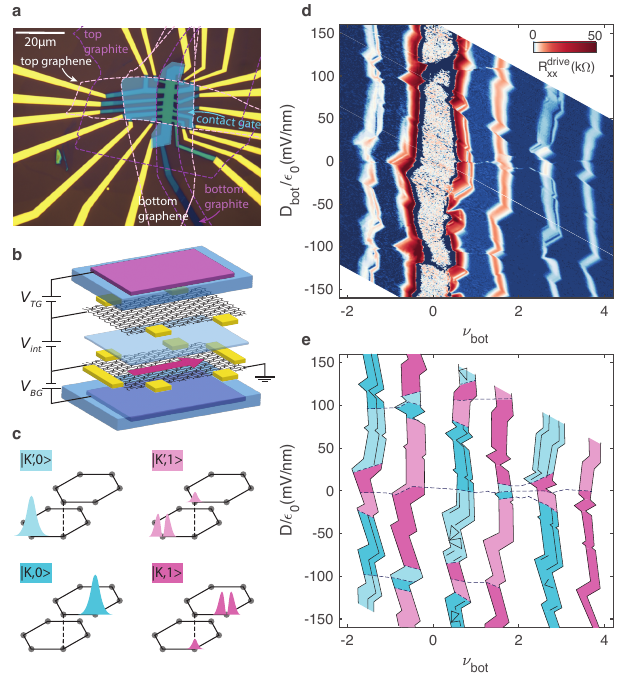}
    \caption{\textbf{a,} Device picture with dashed lines denoting the original flake placement. The semi-transparent region is covered by contact gates. \textbf{b,} Schematic of the double layer structure and device configuration. \textbf{c,} Illustration of valley and orbital flavor of the zero-energy Landau levels in BLG. Spin flavor is not included. Valley flavor is tied with layer polarization. \textbf{d,} Drive resistance $R_{\text{xx}}^{\text{drive}}$ (bottom layer) as a function of displacement field $D_{\text{bot}}$ and filling factor $\nu_{\text{bot}}$. \textbf{e,} Schematic of the main features in \textbf{d}. }
    \label{fig1}
\end{figure*}
\begin{figure*}
    \centering
    \includegraphics{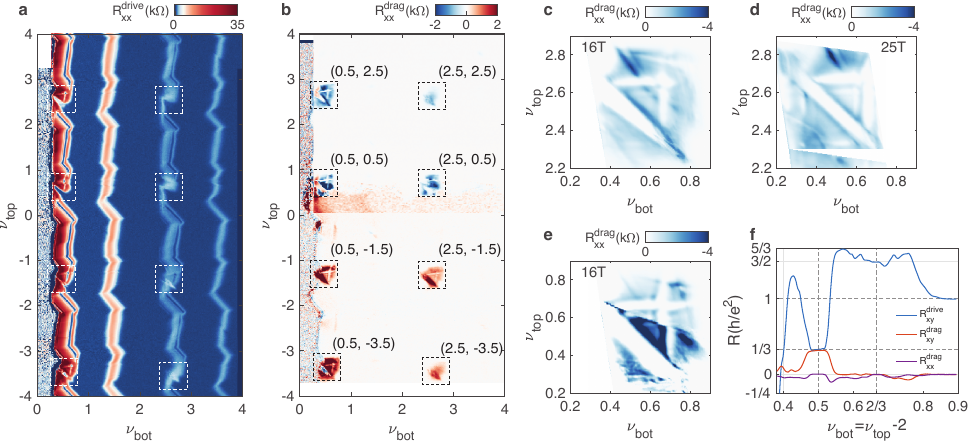}
    \caption{\textbf{a,} $R_{\text{xx}}^{\text{drive}}$ and \textbf{b,} $R_{\text{xx}}^{\text{drag}}$ as a function of $\nu_{\text{top}}$ and $\nu_{\text{bot}}$. \textbf{c,} $R_{\text{xx}}^{\text{drag}}$ near $(\nu_{\text{top}}, \nu_{\text{bot}})=(0.5, 2.5)$ at $B = 16$~T. The colormap limits are chosen to highlight drag resistance and suppress spurious signals due to contact quality.\textbf{d,} same as \textbf{c} but with $B = 25$~T. \textbf{e,} $R_{\text{xx}}^{\text{drag}}$ near $(\nu_{\text{top}}, \nu_{\text{bot}})=(0.5, 0.5)$ at $B = 16$~T. \textbf{f,} Line cuts across the EC state in \textbf{c} showing $R_{\text{xx}}^{\text{drag}}$, $R_{\text{xy}}^{\text{drag}}$ and $R_{\text{xy}}^{\text{drive}}$.}
    \label{fig2}
\end{figure*}
\begin{figure*}
    \centering
    \includegraphics{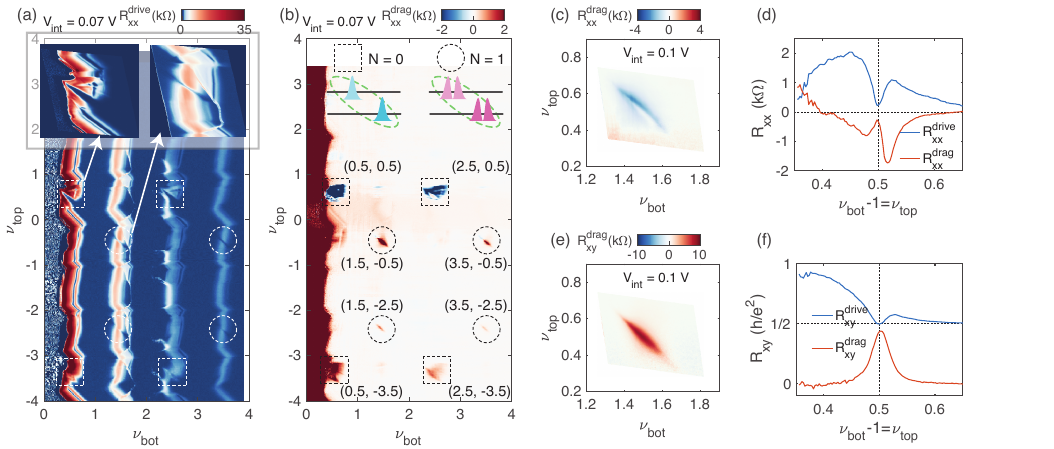}
    \caption{\textbf{a,} $R_{\text{xx}}^{\text{drive}}$ as a function of $\nu_{\text{top}}$ and $\nu_{\text{bot}}$ at $V_{\text{int}}=0.07$~V. Dashed squares mark interlayer states formed by N = 0 orbitals. Dashed circles mark interlayer states formed by $N = 1$ orbitals. \textbf{b,} The corresponding drag signal $R_{\text{xx}}^{\text{drag}}$. The top insets show cartoons illustrating the two types of interlayer states formed by coupling \(N=\) 0 and 1 LL orbitals. \textbf{c,} $R_{\text{xx}}^{\text{drag}}$ and \textbf{e,} $R_{\text{xy}}^{\text{drag}}$ near $(\nu_{\text{top}}, \nu_{\text{bot}})=(1.5, 0.5)$ at $V_{\text{int}}=0.1$~V. \textbf{d} and \textbf{f,} Line cuts across the state showing all other transport channels. Dashed lines denote the state location and quantization value.}
    \label{fig3}
\end{figure*}
\begin{figure*}
    \centering
    \includegraphics{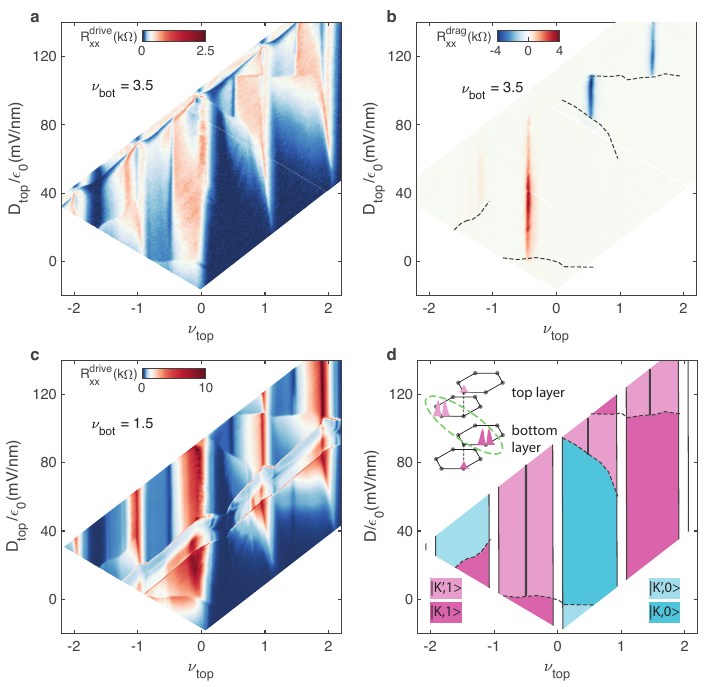}
    \caption{\textbf{a,} $R_{\text{xx}}^{\text{drive}}$ at fixed $\nu_{\text{bot}}=3.5$ as a function $D_{\text{top}}$ and $\nu_{\text{top}}$. \textbf{b,} $R_{\text{xx}}^{\text{drag}}$ in the same range, showing clearly interlayer QH states. \textbf{c,} $R_{\text{xx}}^{\text{drive}}$ at fixed $\nu_{\text{bot}}=1.5$. \textbf{d,} Schematic of the features in the resistance data. This shows the $N=1$ interlayer QH states form only when the second LL component in the $N=1$ orbitals are in proximate layers.}
    \label{fig4}
\end{figure*}

\clearpage
\begin{acknowledgments}
\textbf{Acknowledgment:} The major part of the experiment was supported by the US Department of Energy (DOE) (DE-SC0012260). A portion of this work was performed at the National High Magnetic Field Laboratory, which is supported by National Science Foundation Cooperative Agreement No. DMR-2128556* and the State of Florida. K.W. and T.T. acknowledge support from the JSPS KAKENHI (grants 20H00354 and 23H02052) and World Premier International Research Center Initiative (WPI), MEXT, Japan.
\end{acknowledgments}

\paragraph{}
Z.H. and P.K. conceived the experiment. Z.H. fabricated the devices. Z.H. and A.M.Z. performed the measurements and analysis. K.W and T.T. provided the hBN crystals. P.K. supervised the project. All authors discussed the results and contributed to the manuscript.

\clearpage
\nocite{britnell_electron_2012, lee_electron_2011, gorbachev_strong_2012, Shaowen_competing_2019, PhysRevLett.107.176602, PhysRevX.12.031019, PhysRevLett.132.046603, kumar_quarter_2025, PhysRevLett.122.137701, zhang_excitons_2025, PhysRevB.101.085412}
\bibliography{double_bilayer_paper_reference}
%

%% file: SM_arxiv.tex
\section{S1. Capacitance Model}

Fig.~\ref{fig:SI_schematic}a shows the schematic diagram of the device again. $V_{\text{TG}}, V_{\text{BG}}, V_{\text{int}}$ are voltages applied to the top graphite gate, bottom graphite gate, and top bilayer graphene (BLG), respectively, with the bottom BLG grounded. Taking $V_{\text{est}}$, $V_{\text{esb}}$ and $V_{\text{esint}}$ as electrostatic potentials for the top, bottom, and interlayer capacitors, respectively, we have the following equilibrium equalities.
    \[
    V_{\text{TG}}-V_{\text{int}} = V_{\text{est}} + \mu_t
    \]
    \[
    V_{\text{BG}} = V_{\text{esb}} + \mu_b
    \]
    \[
    V_{\text{int}} = V_{\text{esint}} + \mu_t - \mu_b
    \]
\(\mu_t\) and \(\mu_b\) are chemical potentials of the top and bottom BLG, respectively. The charge accumulation due to each electrostatic voltage term is \(
    q_1 = C_tV_{\text{est}} \), \( q_2 = C_bV_{\text{esb}} \), \( q_3 = C_{\text{int}}V_{\text{esint}}
    \), and $C_t$, $C_{\text{int}}$ and $C_b$ are geometric capacitances between the top graphite and the top BLG, between the top and bottom BLG, and between the bottom BLG and bottom graphite, respectively. Then the general formulas relating the displacement field, carrier density, and chemical potential in each BLG with the gate voltages are as follows:
    \[en_{\text{top}} = q_1+q_3 = C_t(V_{\text{TG}}-V_{\text{int}}-\mu_t)+C_{\text{int}}(-V_{\text{int}}+\mu_b-\mu_t)\] 
    \[en_{\text{bot}} = q_2-q_3 = C_b(V_{\text{BG}}-\mu_b)-C_{\text{int}}(-V_{\text{int}}+\mu_b-\mu_t)\]
    \[D_{\text{top}} = \frac{-q_1+q_3}{2} = \frac{1}{2}(-C_t(V_{\text{TG}}-V_{\text{int}}-\mu_t)+C_{\text{int}}(-V_{\text{int}}+\mu_b-\mu_t))\]
    \[D_{\text{bot}} = \frac{q_2+q_3}{2} = \frac{1}{2}\left(C_b(V_{\text{BG}}-\mu_b)+C_{\text{int}}(-V_{\text{int}}+\mu_b-\mu_t)\right)\]

 Here, $e$ is the electron charge, $\epsilon_0$ is the vacuum permittivity, and $C_{\text{t}}$, $C_{\text{int}}$ and $C_{\text{b}}$ are the geometric capacitances between the top graphite and top BLG, between the top and bottom BLG layers, and between the bottom BLG and bottom graphite, respectively. When determining the rough location in the displacement field-density map, such as those in Fig.~1d and c, Fig.~2a and b, Fig.~3a and b, and Fig.~4a-d, we can often first ignore the chemical potential terms. Then the above formulas reduce to:

\[n_{\text{top}}=\frac{1}{e}\left(C_{\text{t}}(V_{\text{TG}}-V_{\text{int}})-C_{\text{int}}V_{\text{int}}\right)\] \[n_{\text{bot}}=\frac{1}{e}(C_{\text{b}}V_{\text{BG}}+C_{\text{int}}V_{\text{int}})\]
\[D_{\text{top}}/\epsilon_0=\frac{1}{2}(C_{\text{t}}(V_{\text{TG}}-V_{\text{int}})+C_{\text{int}}V_{\text{int}})\]
\[D_{\text{bot}}/\epsilon_0= \frac{1}{2}(-C_{\text{b}}V_{\text{BG}}+C_{\text{int}}V_{\text{int}})\]

\begin{figure}
\centering
\includegraphics[width=0.4\textwidth]{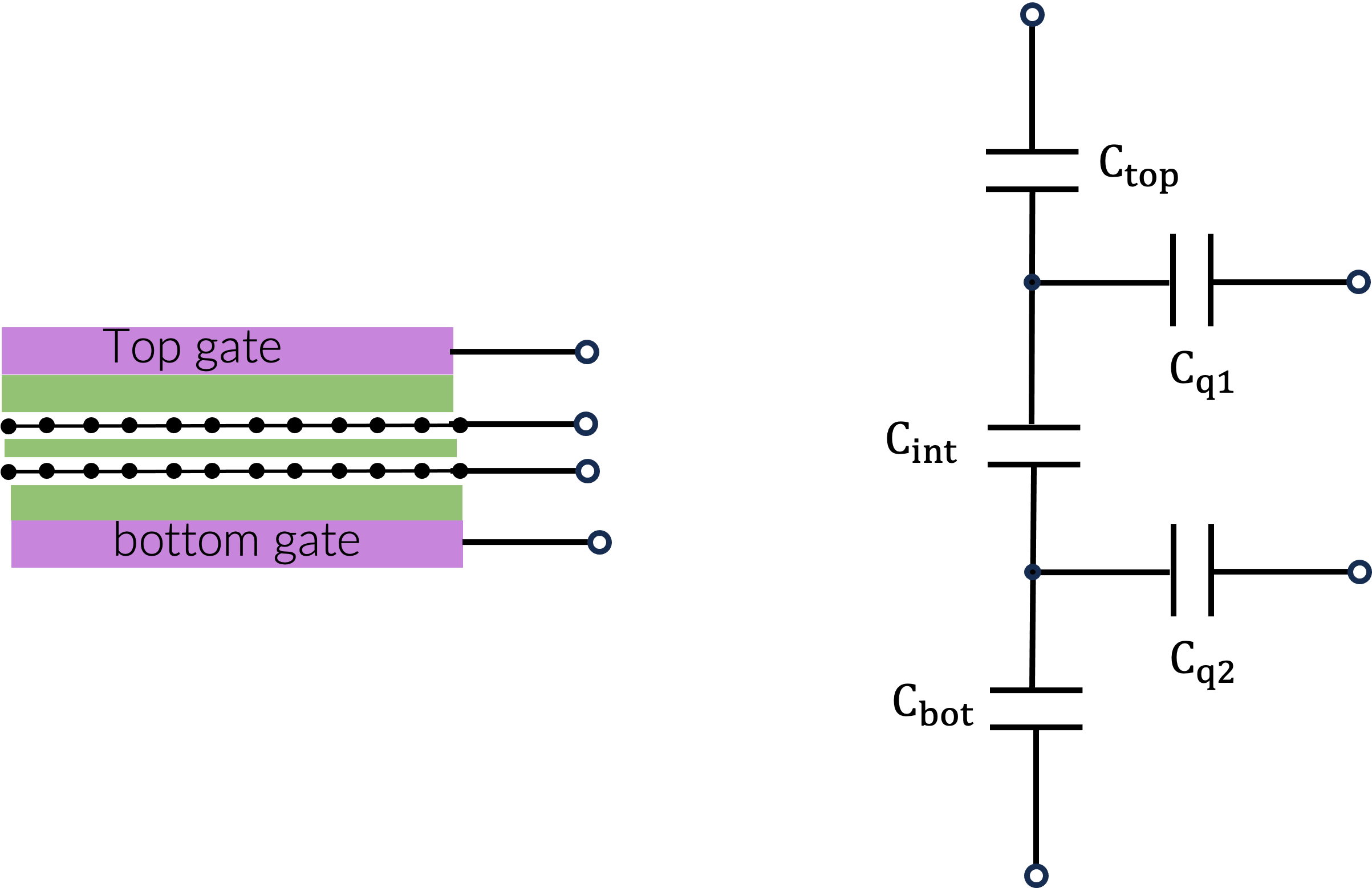}
\caption{\label{fig:SI_double_layer_circuit} \textbf{Left}, schematic of the gating configuration for a double layer graphene device. \textbf{Right}, the corresponding capacitance network circuit model.}
\end{figure}

However, whenever zooming in to specific interlayer state clusters or when an accurate determination of filling factors is needed, such as those in Fig.~2c-f and Fig.~3c-f, this simplified approach becomes insufficient. In those cases, we instead determine the filling factors by calibrating against experimentally observed features in the data. This procedure follows the established practice in previous studies (see for example, the Supplementary Information of \cite{li_pairing_2019}), and we describe the method again in detail below. 

The chemical potential contribution to the filling factor is often incorporated through a  quantum capacitance, \(C_q=e^2\rho\), where \(e\) is electron charge and \(\rho\) is the density of states. Then our double layer system can be modeled as a capacitance network circuit shown in Fig. \ref{fig:SI_double_layer_circuit} with the 4 open ports corresponding to the top and bottom gates and two BLG layers.

Here \(C_{q1}\) and \(C_{q2}\) are quantum capacitances of the top and bottom BLG, respectively. When the top or bottom BLG has a high density of states and can screen efficiently, its quantum capacitance becomes much larger than the geometric capacitance. In this limit, the quantum capacitance has negligible impact and the carrier densities are solely determined by geometric capacitances, thus the bottom gate does not affect the filling factor in the top BLG due to the perfect screening of the bottom BLG, and similarly no top gate tuning for the bottom BLG. 
\begin{figure}
\centering
\includegraphics[width=0.6\textwidth]{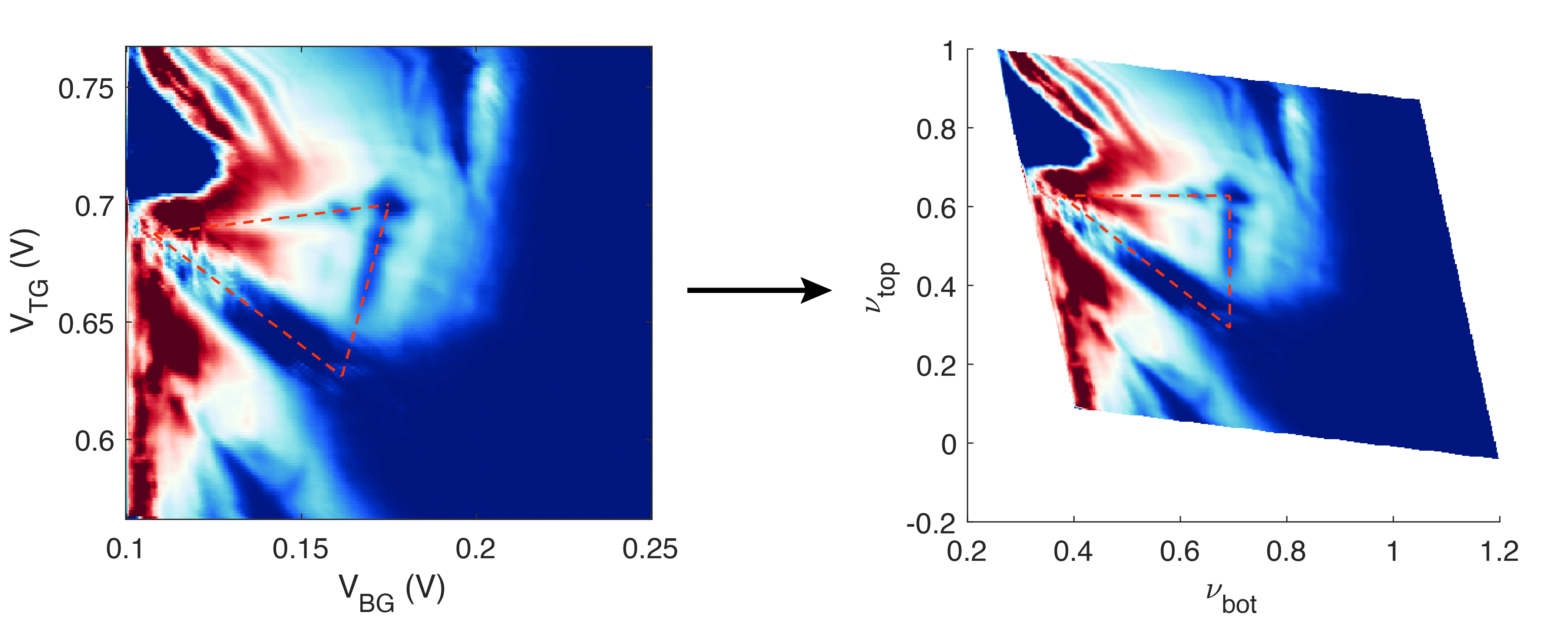}
\caption{\label{fig:SI_zoom2D_convert} Zoom-in resistance map as a function of gates converted to top and bottom layer filling factors.}
\end{figure}

In contrast, when both BLGs are in partially filled LL, their densities of states–and hence their screening capabilities–are finite. Under these conditions, neither layer can fully screen electric fields, allowing both gates to influence the carrier densities and filling factors of both BLG layers. More specifically, if we define capacitance matrix \(C_{ij}=n_i/V_j\)~\cite{zibrov_even-denominator_2017}, the effect of the bottom gate voltage on the top BLG carrier density can then be expressed \(C_{\text{tB}}=n_{\text{top}}/V_{\text{BG}}=C_{\text{int}}C_{\text{bot}}/(C_{\text{int}}+C_{\text{bot}}+\rho_{\text{bot}})\). If we assume, within the same LL, the screening capability represented by the density of states remains approximately  constant, then the gate-density relation can be simplified. This is a reasonable assumption because the energy gaps of 2-component FQH states are mostly below 5K and the variation due to energy gaps within a LL is a small perturbation~\cite{li_pairing_2019}. Under this assumption, the relation between the carrier densities and top and bottom gates (say, when \(V_{\text{top}}\) is fixed) is reduced to a linear transformation: \(n_{\text{top}} = C_{\text{top}}V_{\text{TG}}+C_{\text{tB}}V_{\text{BG}}\),  \(n_{\text{bot}} = C_{\text{bot}}V_{\text{BG}}+C_{\text{bT}}V_{\text{TG}}\). In our data, as illustrated in Fig. \ref{fig:SI_zoom2D_convert} for \((\nu_{\text{bot}}, \nu_{\text{top}})\sim\)(0.5, 2.5), we can calibrate off the individual intralayer 2/3 FQH states so that they show as horizontal or vertical features at constant fillings in \(\nu_{\text{bot}}-\nu_{\text{top}}\) maps.

\begin{figure}
\centering
\includegraphics{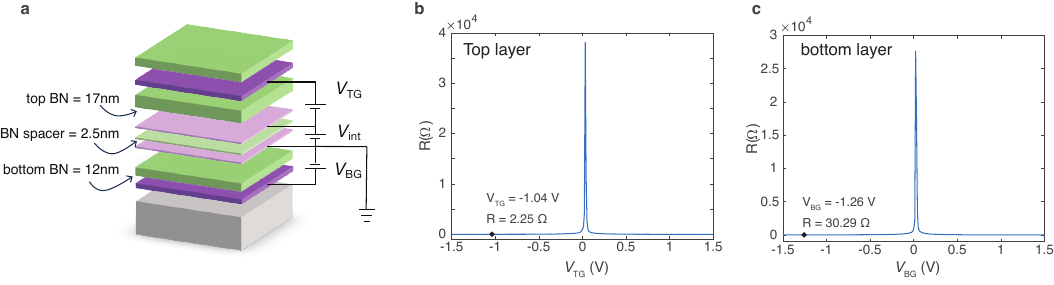}
\caption{\label{fig:SI_schematic} \textbf{a,} device schematic showing thickness of the BN layers and gate configurations. \textbf{b} and \textbf{c,} resistance at zero magnetic field in the top and bottom layer respectively. The sharp charge neutrality peaks show the high quality of both of the two Bernal bilayer graphene layers.}
\end{figure}

\begin{figure}
\centering
\includegraphics{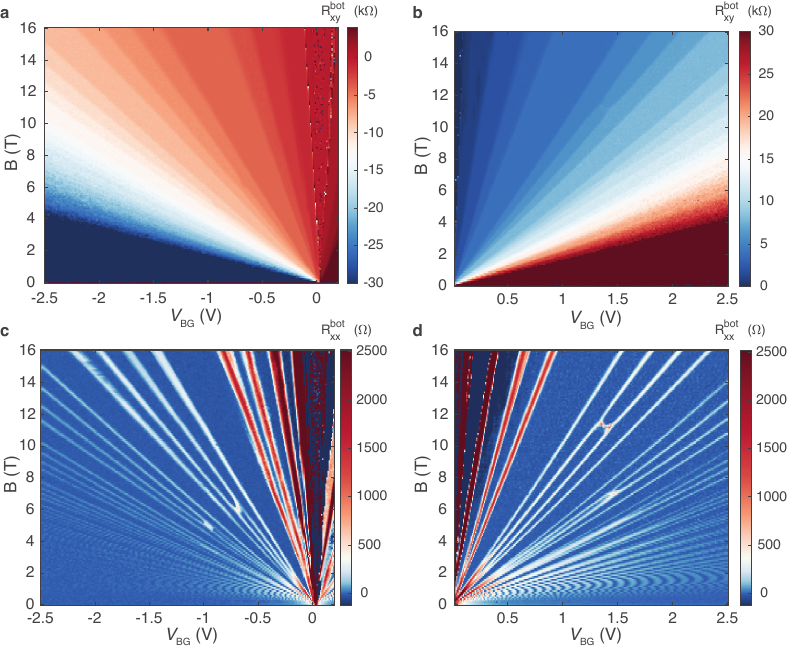}
\caption{\label{fig:SI_fans} Landau fan diagrams of the bottom layer as a function of \(V_\text{BG}\). \textbf{a} and \textbf{b}, Hall resistance \(R_{\text{xy}}\) on the negative (hole-doped) and positive (electron-doped) \(V_\text{BG}\) side. \textbf{c} and \textbf{d}, longitudinal resistance \(R_{\text{xx}}\). }
\end{figure}

\section{S2. Accurate Extraction of BN thickness}
\begin{figure}
\centering
\includegraphics{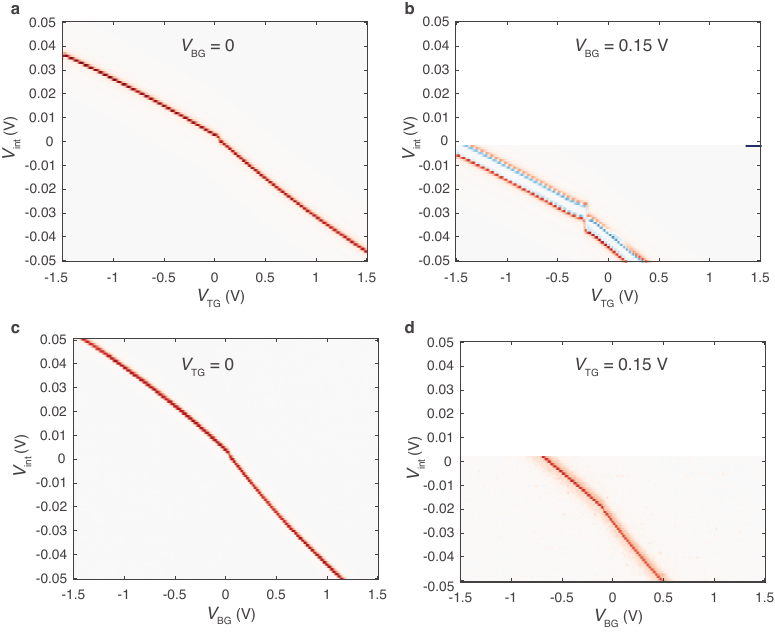}
\caption{\label{fig:SI_chemical} BN spacer layer thickness extraction. \textbf{a} and \textbf{b,} \(R_{\text{xx}}\) in the top bilayer graphene as a function of \(V_{\text{int}}\) and \(V_{\text{TG}}\), with \(V_{\text{BG}}\) fixed to 0 and 0.15~V, respectively. \textbf{c} and \textbf{d,} \(R_{\text{xx}}\) in the bottom bilayer graphene as a function of \(V_{\text{int}}\) and \(V_{\text{BG}}\), with \(V_{\text{TG}}\) fixed to 0 and 0.15~V, respectively.}
\end{figure}

Figs.~\ref{fig:SI_schematic}b and c show the charge neutrality resistance peaks (CNPs) in the two BLG layers, respectively, where we measure their longitudinal resistance \(R_{\text{xx}}\) separately using only the top or bottom gate. The sharp peaks demonstrate the high quality of the individual BLG samples. Fig.~\ref{fig:SI_fans} shows transverse and longitudinal resistance \(R_{\text{xy}}\) and \(R_{\text{xx}}\) in the bottom layer. Clean data with good quantization in \(R_{\text{xy}}\) and the corresponding zero \(R_{\text{xx}}\) again demonstrate the high quality of the sample. 

When converting the gate voltages to \(D\) or \(n\) associated with the BLG layers, it is important to know the thicknesses of the BN dielectric layers well. Although atomic force microscopy (AFM) can reliably measure relatively thick BN layers, giving the thickness of the top BN (between the top gate and top BLG) \(\approx\)17~nm and bottom BN (between the bottom gate and bottom BLG) \(\approx\)12~nm, it can be difficult to accurately measure a thin BN layer. In our experiment, we utilize the principle of measuring the chemical potential of one graphene layer using an adjacent graphene layer to extract the thickness of the BN spacer layer between layers. 

Using the formulas we derived above and setting \(n_{\text{bot}}=0\) and \(\mu_b=0\), corresponding to staying at the charge neutrality of the bottom BLG, we obtain the following.
\[en_{\text{top}} = C_tV_{\text{TG}}+\frac{C_b(C_t+C_{\text{int}})}{C_{\text{int}}}V_{\text{BG}}\]
\[\mu_t = -V_{\text{int}}-\frac{C_b}{C_{\text{int}}}V_{\text{BG}}\]

These formulas mean that if we follow the CNP in the bottom BLG — for instance, the resistance peak in Fig.~\ref{fig:SI_chemical}a and b — we can directly measure the \(n_{\text{top}}\) and \(\mu_t\) at each point along the peak using the gate voltages at that point and geometric capacitance values. CNP peak in Fig.~\ref{fig:SI_chemical}b is shifted to the lower left compared to Fig.~\ref{fig:SI_chemical}a due to a finite \(V_{\text{BG}}\). Using this shift, we can directly extract \(C_b/C_{\text{int}}\) using the formulas above and obtain an accurate value of the BN spacer layer thickness. Similarly, by tracking the CNP in the top layer, as shown in Fig.~\ref{fig:SI_chemical}a and b, we can also extract \(C_t/C_{\text{int}}\), allowing a consistent check of the relative thicknesses between the BN layers.

The hBN spacer thickness is deliberately chosen to minimize the ratio \(d/l_B\) to enhance interlayer coupling while simultaneously suppressing interlayer tunneling. Our choice of \(d\) = 2.5 nm hBN spacer, corresponding to a \(d/l_B\) ~ 0.4 at B = 16 T, is consistent with the previous studies. For example, Liu et al. \cite{liu_interlayer_2019} employed a 2.5nm thick hBN spacer and J. I. A. Li et al. \cite{li_pairing_2019} used a 2.7 nm spacer. In both cases, strong interlayer coupling could be achieved, as evidenced by the plethora of interlayer FQH states reported in these papers. 

\begin{figure}
\centering
\includegraphics[width=0.4\textwidth]{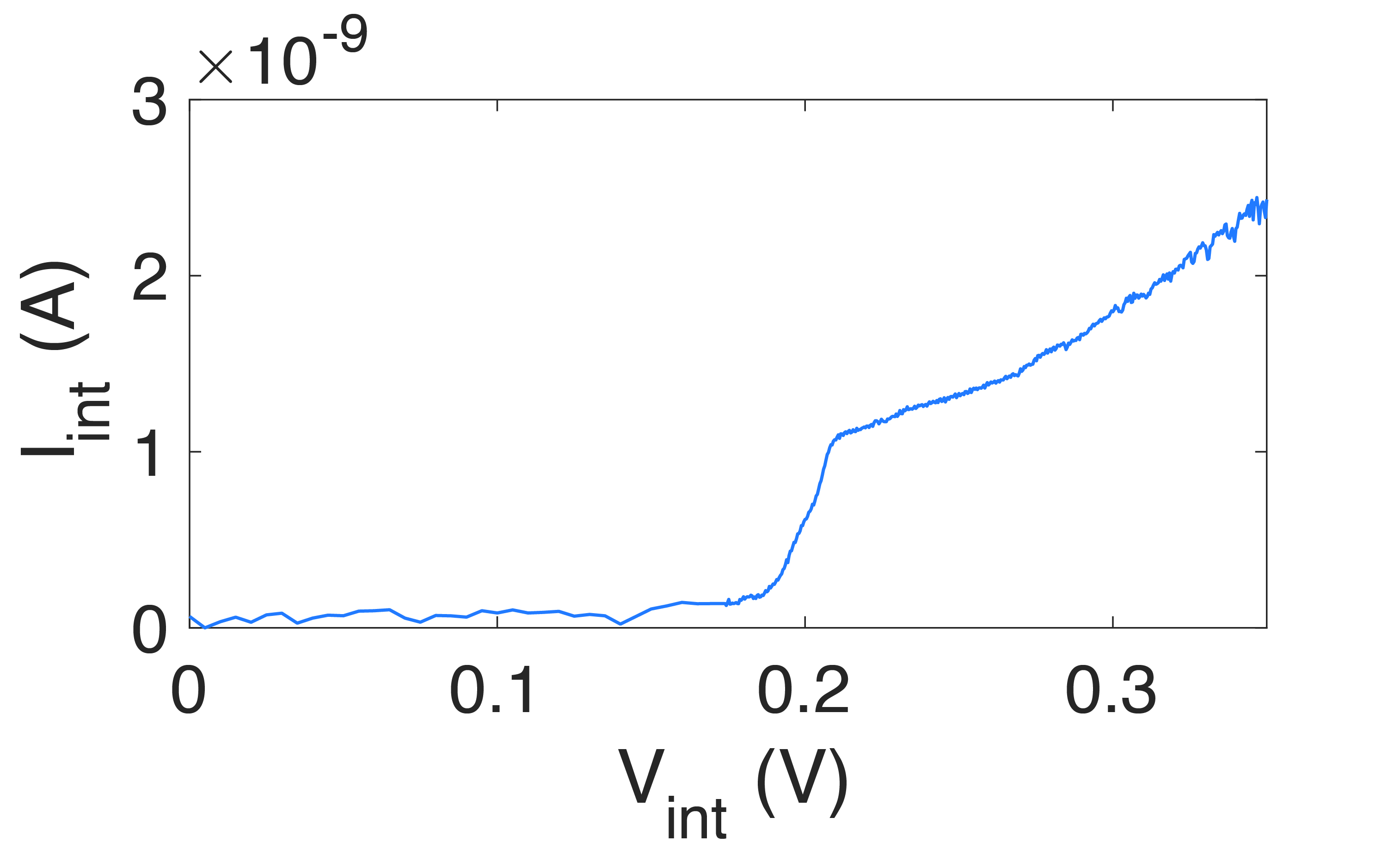}
\caption{\label{fig:SI_leakage} Leakage current \(I_{\text{int}}\) through the hBN spacer as a function of interlayer bias voltage \(V_{\text{int}}\)}.
\end{figure}

On the other hand, previous tunneling studies have shown that for a hBN spacer with d \(>\) 1.5 nm at small biases (\(<V_{\text{int}}\)~ 1 V), interlayer tunneling current is negligible. This suppression comes from the large hBN bandgap and lattice orientation mismatch between graphene layers \cite{britnell_electron_2012,lee_electron_2011,gorbachev_strong_2012}. In our experiments, we further suppress tunneling by intentionally misaligning the two bilayer graphene sheets using their natural straight edges. Owing to these considerations, we observe negligible interlayer current for \(V_{\text{int}}<\) 0.2 V (see Fig. \ref{fig:SI_leakage}), which is well above the interlayer bias voltage ranges we explored in the main text.

\section{S3. Additional discussion on our measured BLG QH phase diagram and its possible connection to the absence of interlayer FQH states in N = 1 orbital}

In the main text, we discussed that the presence or absence of the denominator-3 FQH state features provides a visual distinction can help us roughly identify the orbital number \(N=\) 0 or 1 for each region in the $D$ versus $\nu$ phase diagram. However, more accurate identification of the orbital number relies on calculating \(D\) and \(\nu\), and using the QH phase diagram of single BLG as a reference, from which the flavors of both layers can be deduced.

\begin{figure}
\centering
\includegraphics[width=0.7\textwidth]{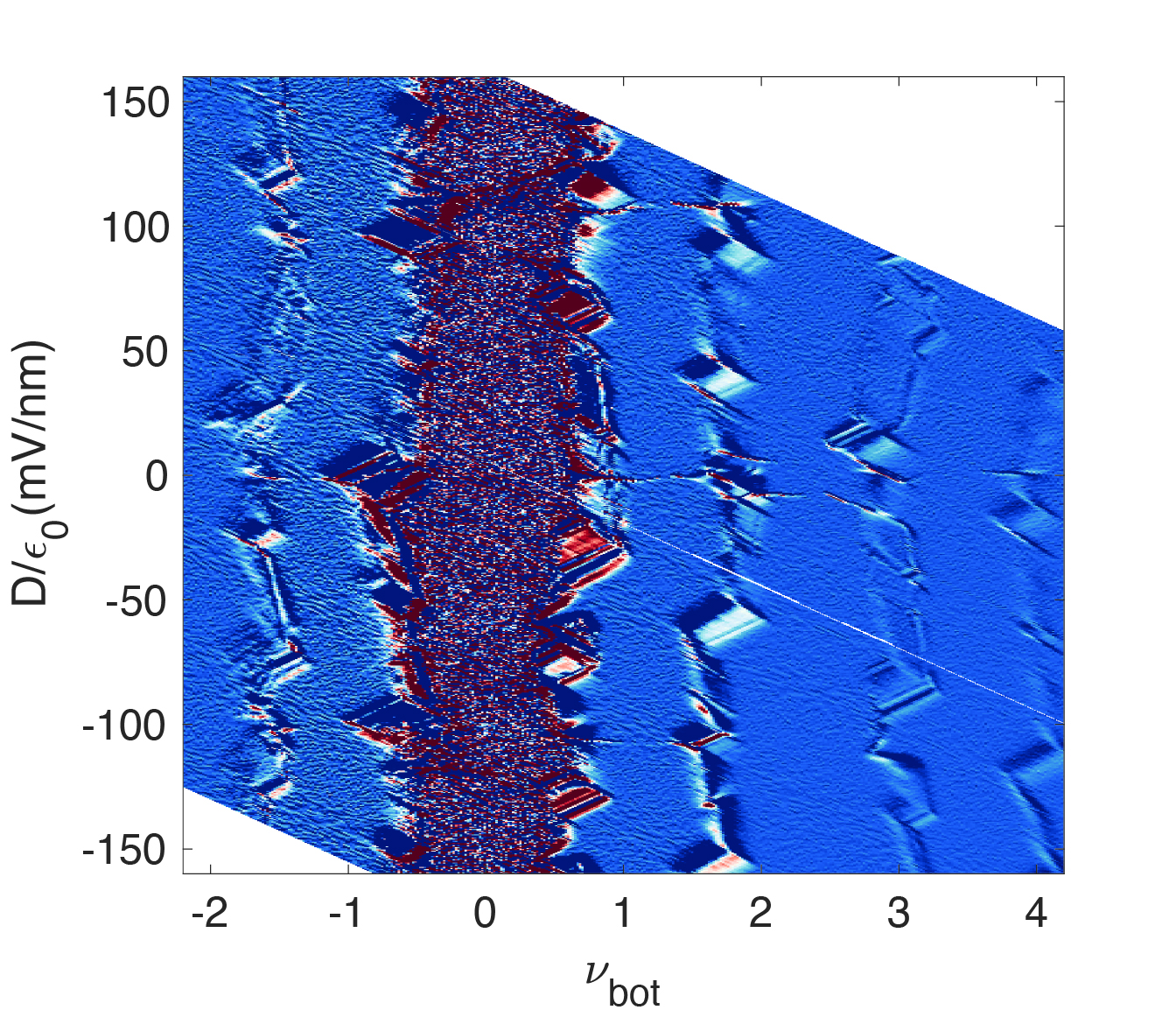}
\caption{\label{fig:SI_derivative} Resistance derivative \(dR_{\text{xx}}^{\text{drive}}/d\nu_{\text{bot}}\) corresponding to the data in Fig.~1d, more clearly showing quantum Hall state and hierarchy transition features.}
\end{figure}

Fig.~\ref{fig:SI_derivative} is presented from the same data as the Fig.~1d, but showing the numerical derivative \(dR_{\text{xx}}^{\text{drive}}/d\nu_{\text{bot}}\) to make the features, especially the flavor transition features more prominent. 


Unlike the N = 0 LL in double mono- or bilayer graphene, in our experiment we don't observe any interlayer FQH states for N = 1 LL. Here are some considerations and what could be done in the future to explore this direction. 

The first factor that stands out is screening from nearby gates. As discussed in the manuscript, the top and bottom hBN dielectric layers have a thickness of 12 nm and 17 nm, respectively. At B = 16 T, the ratio between 12 nm and the magnetic length \(l_B\) is only 1.88. This relatively small separation from the gates significantly enhances screening, leading to substantial modification of the Coulomb interaction scale \(e^2/l_B\). For example, previous studies have shown that proximity to gates can reduce the energy gaps of FQH states~\cite{Shaowen_competing_2019}, and alter the relative strengths of different Haldane pseudopotentials~\cite{PhysRevLett.107.176602}, thereby affecting  the stability of certain FQH states, such as the Moore-Read state. 

\begin{figure}
\centering
\includegraphics[width=0.5\textwidth]{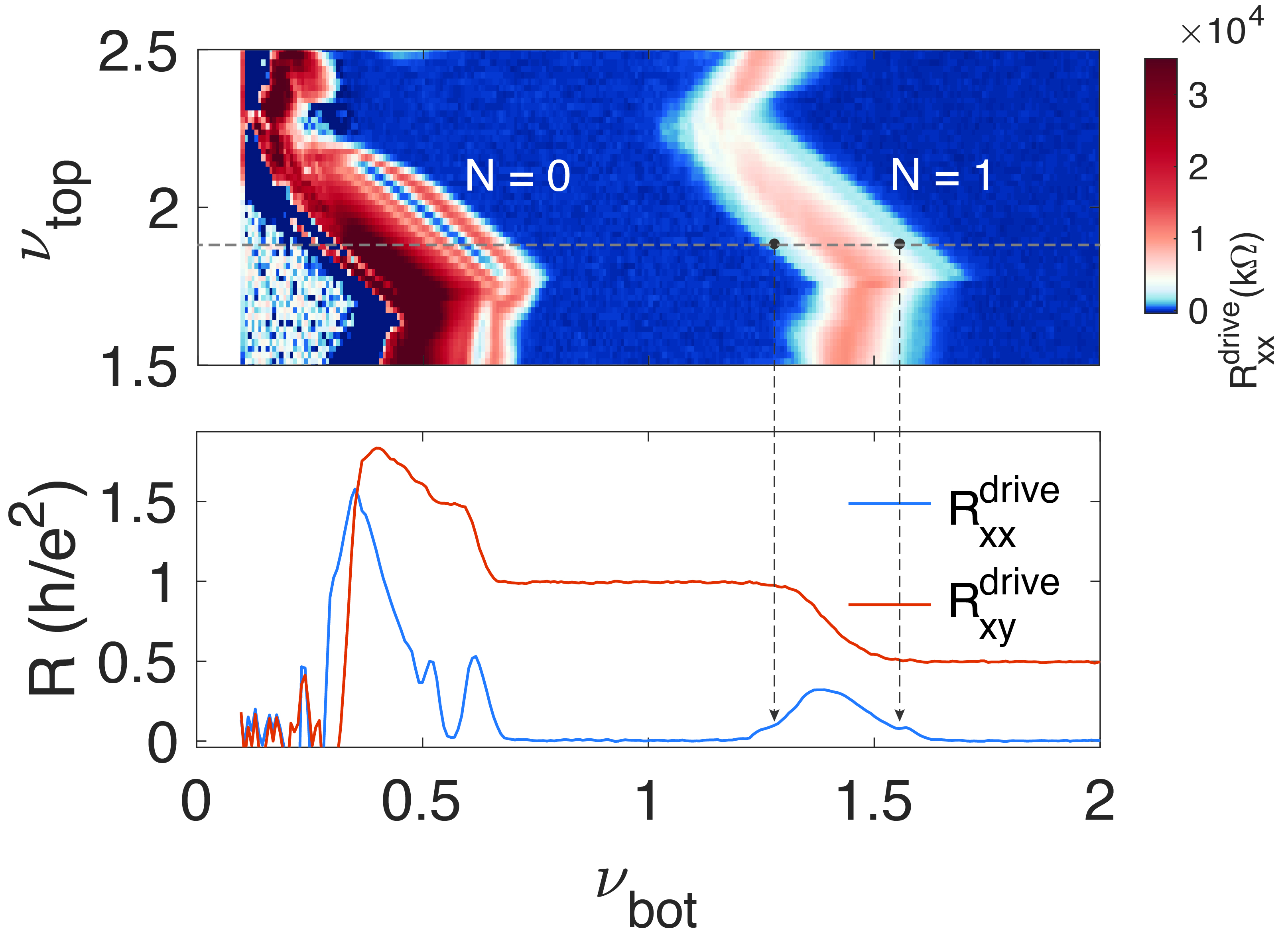}
\caption{\label{fig:SI_N=1_FQH_states} \textbf{top}: bottom layer (drive layer) \(R_{\text{xx}}^{\text{drive}}\) as a function of \(\nu_{\text{bot}}\) and \(\nu_{\text{top}}\) at \(V_{\text{int}}=0.1\)~V, focusing on the contrasting appearances between N = 0 and 1 LL in the bottom layer. \textbf{bottom}: line cuts of \(R_{\text{xx}}^{\text{drive}}\) and \(R_{\text{xy}}^{\text{drive}}\) at a fixed \(\nu_{\text{top}}\). Dashed arrows mark the weak development of 4/3 and 5/3 FQH states in the N = 1 LL. }
\end{figure}

In our experiment, for N = 1 Landau level (LL) orbitals, we do not observe intralayer even-denominator states in an individual BLG, and only very weak signatures of odd-denominator FQH states. The suppression of intralayer FQH remains true even when the other layer is well inside an integer quantum Hall gap, where screening from that layer should be minimal. An example at \(V_{\text{int}}=0.1\)~V is shown in Fig.~\ref{fig:SI_N=1_FQH_states}:  the top layer is pinned at top=2 integer quantum Hall state, while the bottom layer sweeps through \(0<\nu_{\text{bot}}<2\). The dashed arrows indicate weakly developed 4/3 and 5/3 FQH states, with no observable signatures of the 3/2 state.

Meanwhile, the majority of quantum Hall studies on BLG that report well-developed even- and odd- denominator FQH states have much thicker hBN dielectric layers. For example: Ke Huang, Jun Zhu, et al. \cite{PhysRevX.12.031019} reports devices with hBN layers all greater than 23 nm; Zibrov, Young, et al., \cite{zibrov_even-denominator_2017} 40-50nm BN for all devices; Assouline, Young, et al., \cite{PhysRevLett.132.046603} 48.9/37.2 nm; Kumar, Ronen, et al., \cite{kumar_quarter_2025} all around 30–40nm. 

This comparison suggests that enhanced gate screening in our device, arising from the relatively thin hBN layers, may be responsible for the absence of intralayer and presumably also interlayer FQH states. Future devices incorporating thicker BN dielectrics will be essential to mitigate screening effects and provide better stability in these correlated states.
The second contributing factor could be related to device geometry. In the double layer structure of monolayer graphene, improvements in sample quality, which can be enabled by double graphite gates, were crucial in advancing from observing only interlayer quantum Hall states \cite{liu_quantum_2017, li_excitonic_2017} to resolving more intricate interlayer FQH states \cite{liu_interlayer_2019, li_pairing_2019}. Our device is already encapsulated by top and bottom graphite gates, the sample quality therefore is less likely a limiting factor. 

Nevertheless, even for devices of comparable quality, Corbino geometry often exhibits a larger number of interlayer FQH states with significantly improved visibility \cite{li_pairing_2019}. This is likely because the Corbino geometry eliminates edge transport, which depends sensitively on equilibrium between contacts and edge modes, and is also susceptible to more disordered physical device edges. By suppressing edge contribution, the Corbino devices provide better sensitivity to bulk incompressible states \cite{PhysRevLett.122.137701}. Therefore, in our current double layer structure of Bernal bilayer graphene, it is possible that additional more fragile interlayer FQH states are present but remain unresolved due to the limited probing sensitivity. Such states, if they exist, can be accessible using more bulk sensitive geometries or probes like Corbino devices. 
Another important point we wish to emphasize is that a recent Corbino-geometry study of double monolayer graphene~\cite{zhang_excitons_2025} revealed a remarkable level of complexity for the interlayer FQH states, comparable to–if not more–the Jain sequence of intralayer FQH states in a monolayer graphene. This result suggests that there is no intrinsic limitation to realizing interlayer FQH states between N = 1 orbitals in double layer BLG, particularly given the robust observations of intralayer FQH states in single BLG layers.

It’s also possible that N = 1 LL intrinsically supports a different, and potentially weaker, interlayer FQH hierarchy structure, in which case further theory exploration is needed. As discussed in the manuscript, the LL index labels orbitals with different wavefunctions, which directly modify the Coulomb interaction and consequently, the stability and nature of the resulting FQH states. For example, \cite{PhysRevB.101.085412} presents a theoretical study of interlayer FQH states between N = 0 orbitals, while the theory in Supplementary Information of \cite{shi_bilayer_2022} examines interlayer integer states involving higher orbits (N \(>\) 1). Comparable theoretical treatments specifically addressing interlayer FQH between N = 1 orbitals are still lacking. 

Many of the interlayer fractional states are understood as sharing interlayer vortices in the composite fermion picture. Within this framework, strengthening the interlayer correlation is expected to enhance the stability of interlayer quantum Hall states. This consideration motivates future experiments employing devices with an even thinner hBN spacer to increase interlayer coupling.

\section{S4. Additional Data of Interlayer quantum Hall States}

In the large 2D $\nu_{\text{bot}}-\nu_{\text{top}}$ maps, for instance, Fig~2a and 2b, although the two axes of this diagram, $\nu_{\text{bot}}$ and $\nu_{\text{top}}$ are nominally proportional to $V_{\text{BG}}$ and $V_{\text{TG}}$ when $V_{\text{int}}=0$, the more accurate formula of $\nu_{\text{bot}} = n_{\text{bot}}/n_0$ includes chemical potentials: $n_{\text{bot}}=\frac{1}{e}\left(C_{\text{b}}V_{\text{BG}}-(C_{b}+C_{\text{int}})\mu_{\text{bot}}+C_{\text{int}}\mu_{\text{top}}\right)$, where $\mu_{\text{top}}$ and $\mu_{\text{bot}}$ are the chemical potentials of top and bottom BLG layers, respectively. Therefore, if we follow a resistance feature of the bottom BLG that signals fixed $n_{\text{bot}}$ and $\mu_{\text{bot}}$, the $V_{\text{BG}}$ value of that feature is directly proportional to $\mu_{\text{top}}$. Indeed a negative slope in the $\nu_{\text{bot}}-\nu_{\text{top}}$ plane indicates a gapped state, where the chemical potential of top BLG suddenly shifts

\begin{figure}
\centering
\includegraphics[width=0.9\textwidth]{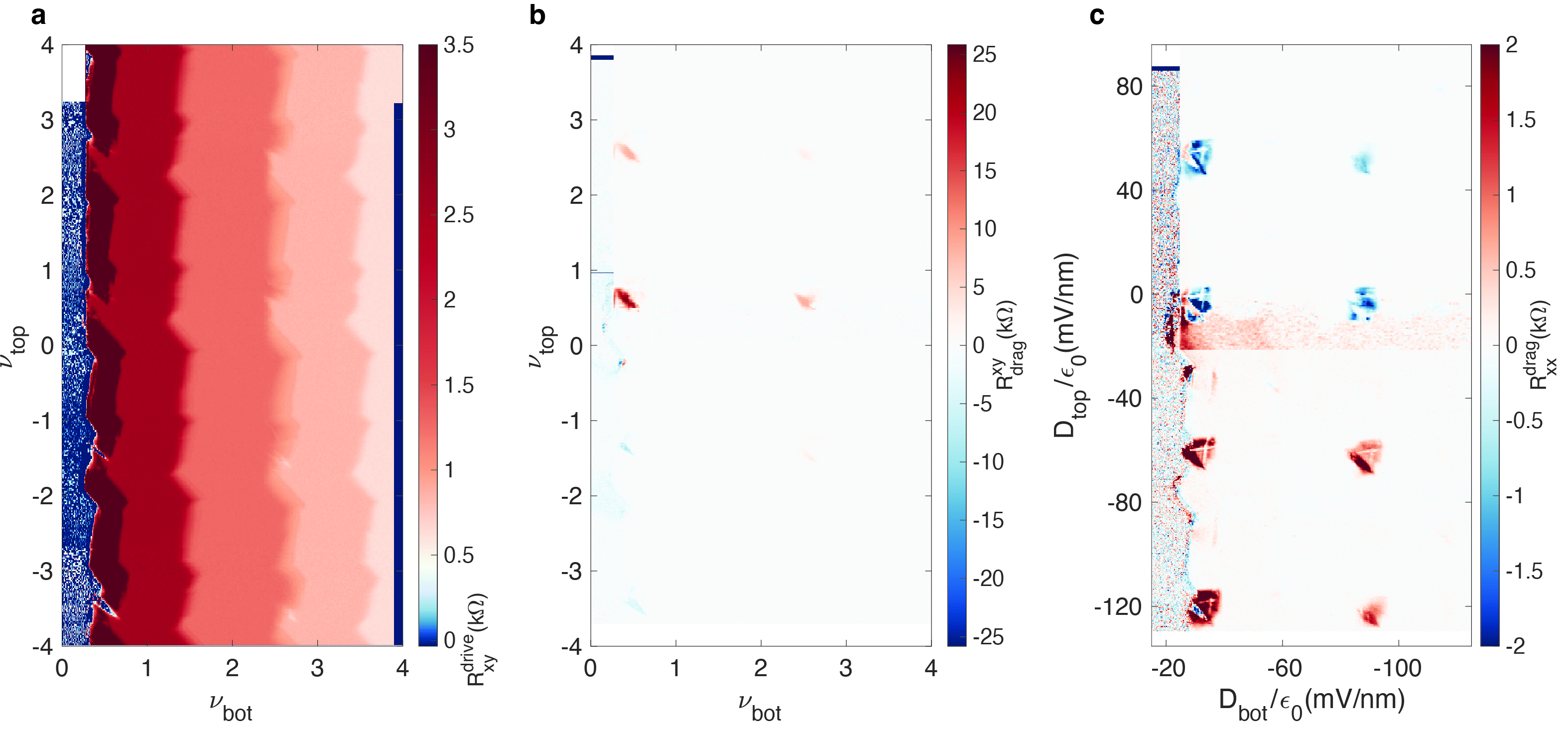}%
\caption{\label{fig:SI_other_channel0} Other resistance channels at \(V_{\text{int}}=0\). \textbf{a,} \(R_{\text{xy}}^{\text{drive}}\). \textbf{b,} \(R_{\text{xy}}^{\text{drag}}\). \textbf{c,} \(R_{\text{xx}}^{\text{drag}}\) plotted as a function of \(D_{\text{top}}\) and \(D_{\text{bot}}\).}
\end{figure}

\begin{figure}
\centering
\includegraphics[width=0.9\textwidth]{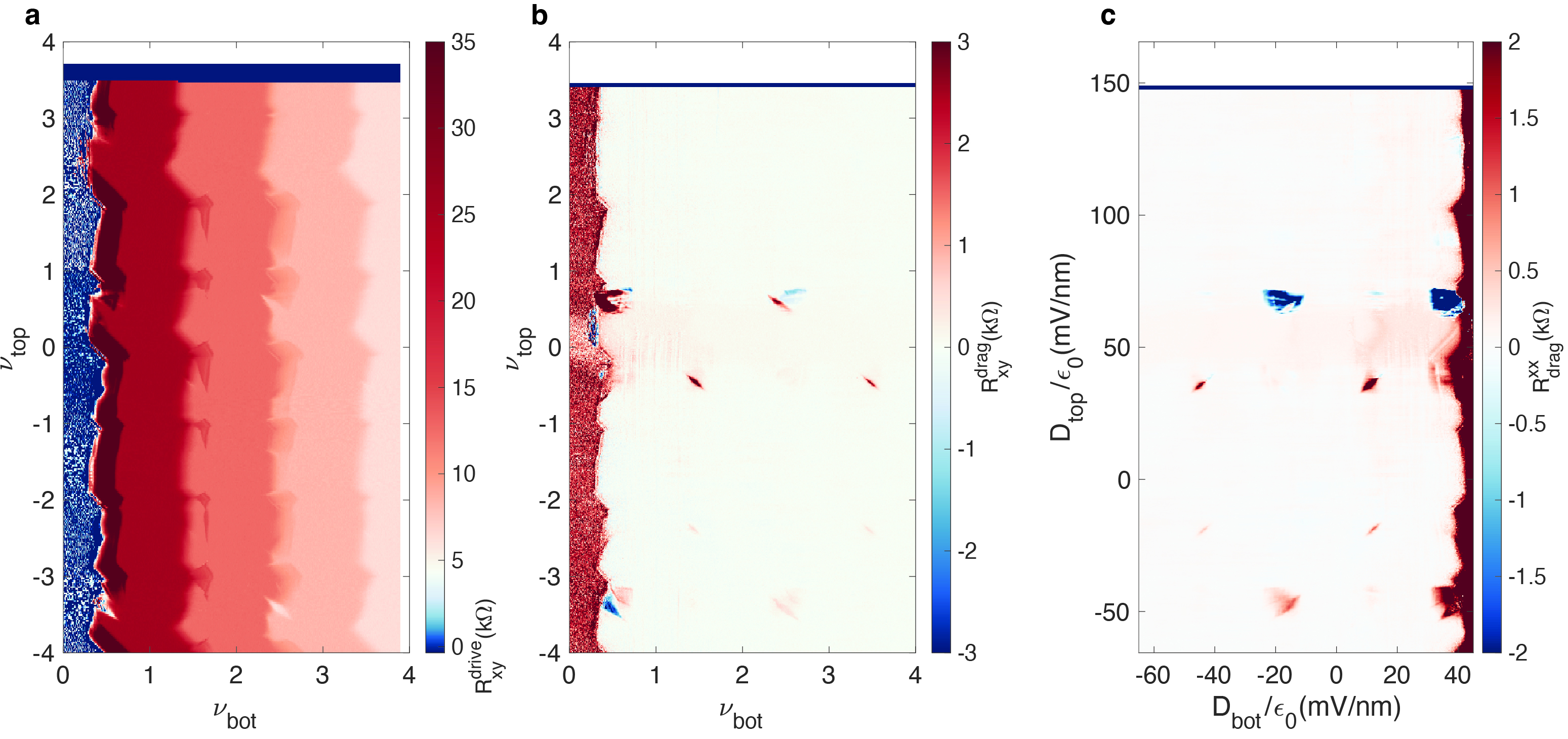}%
\caption{\label{fig:SI_other_channel007} Other resistance channels at \(V_{\text{int}}=0.07~V\). \textbf{a,} \(R_{\text{xy}}^{\text{drive}}\). \textbf{b,} \(R_{\text{xy}}^{\text{drag}}\). \textbf{c,}\(R_{\text{xx}}^{\text{drag}}\) plotted as a function of \(D_{\text{top}}\) and \(D_{\text{bot}}\).}
\end{figure}

Fig.~\ref{fig:SI_other_channel0} (\(V_{\text{int}} =\) 0) and \ref{fig:SI_other_channel007} ( \(V_{\text{int}} =\) 0.07~V) show complementary resistance channels to those presented in Fig.~2 and Fig.~3 in the main text. Importantly, Fig.~\ref{fig:SI_other_channel0}c and \ref{fig:SI_other_channel007}c plot the data as a function of the displacement fields of the top and bottom BLG. By reading off the corresponding (\(D_{\text{bot}}\), \(D_{\text{top}}\)) and (\(\nu_{\text{bot}}\), \(\nu_{\text{top}}\)) values for each interlayer state and consulting the single BLG phase diagram, we can unambiguously identify the flavor and orbital of each BLG underlying the interlayer state. 

\begin{figure}
\centering
\includegraphics[width=0.9\textwidth]{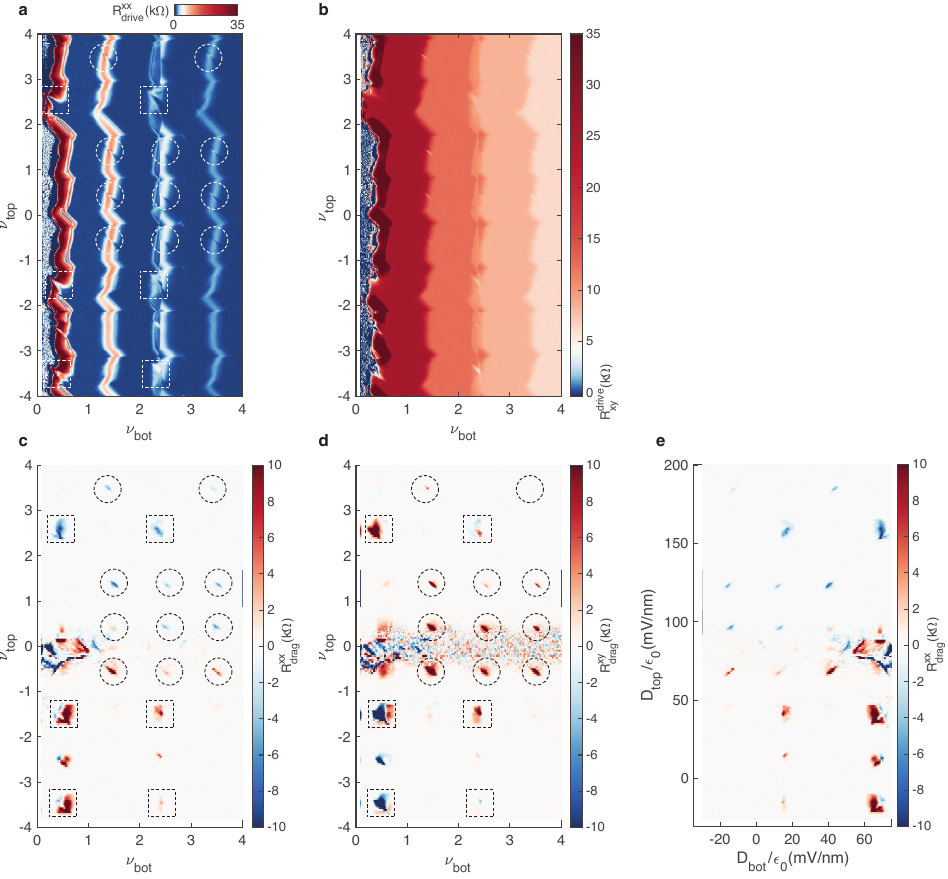}%
\caption{\label{fig:SI_other_channel01} All resistance channels at \(V_{\text{int}}=0.1~V\). \textbf{a,} \(R_{\text{xx}}^{\text{drive}}\). \textbf{b,} \(R_{\text{xy}}^{\text{drive}}\). \textbf{c,} \(R_{\text{xx}}^{\text{drag}}\). \textbf{d,} \(R_{\text{xy}}^{\text{drag}}\). \textbf{e,} \(R_{\text{xx}}^{\text{drag}}\) plotted as a function of \(D_{\text{top}}\) and \(D_{\text{bot}}\).}
\label{SI_fig4}
\end{figure}

Fig.~\ref{fig:SI_other_channel01} shows all channels at \(V_{\text{int}}=0.1\)~V, in which more N = 1 EC states emerge, marked by dashed circles.

\section{S5. Additional zoomed-in views of interlayer states and semi-quantized state analysis}

\begin{figure}
\centering
\includegraphics[width=0.8\textwidth]{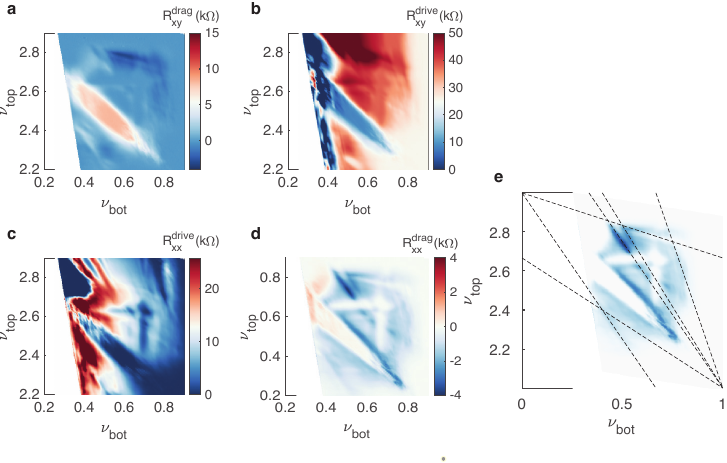}%
\caption{\label{fig:SI_16T_0525} Zoomed-in scans on the N = 0 interlayer states near \((\nu_{\text{bot}}, \nu_{\text{top}}) = (0.5, 2.5)\) at \(V_{\text{int}}=0\) and at 16~T. \textbf{a,} \(R_{\text{xy}}^{\text{drag}}\). \textbf{b,} \(R_{\text{xy}}^{\text{drive}}\). \textbf{c,} \(R_{\text{xx}}^{\text{drive}}\). \textbf{d,} \(R_{\text{xx}}^{\text{drag}}\), the same data as Fig.~2c in the main text. \textbf{e,} Dashed lines overlaid on top of the \(R_{\text{xx}}^{\text{drag}}\) data, marking interlayer fractional states, including $\nu_{\text{bot}}+\frac{2(\nu_{\text{top}}-2)}{3}=1$, $\nu_{\text{bot}}+\frac{3(\nu_{\text{top}}-2)}{5}=1$ and $\frac{\nu_{\text{bot}}}{3}+(\nu_{\text{top}}-2)=1$. There are subtle but less conclusive signals along $\nu_{\text{bot}}+\frac{(\nu_{\text{top}}-2)}{3}=1$, $\nu_{\text{bot}}+\frac{3(\nu_{\text{top}}-2)}{2}=1$ and $\frac{3\nu_{\text{bot}}}{2}+(\nu_{\text{top}}-2)=1$.
}
\end{figure}

\begin{figure}
\centering
\includegraphics[]{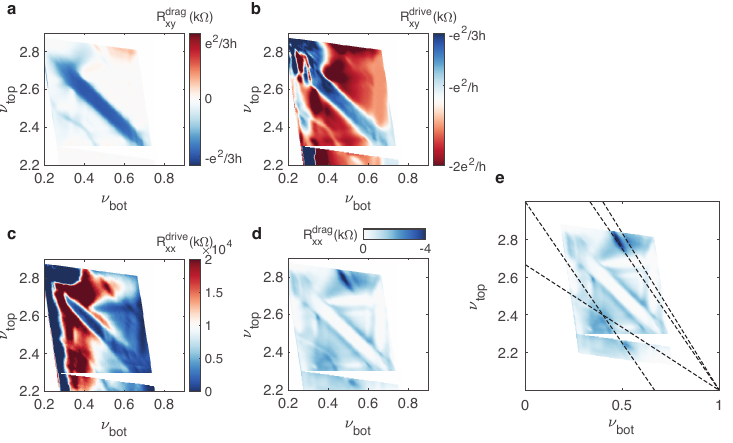}%
\caption{\label{fig:SI_25T_0525} Zoomed-in scans on the N = 0 interlayer states near \((\nu_{\text{bot}}, \nu_{\text{top}}) = (0.5, 2.5)\) at \(V_{\text{int}}=0\) and at 25~T. \textbf{a,} \(R_{\text{xy}}^{\text{drag}}\). \textbf{b,} \(R_{\text{xy}}^{\text{drive}}\). \textbf{c,} \(R_{\text{xx}}^{\text{drive}}\). \textbf{d,} \(R_{\text{xx}}^{\text{drag}}\), the same data as Fig.~2d in the main text. \textbf{e,} dashed lines overlaid on top of the \(R_{\text{xx}}^{\text{drag}}\) data, marking interlayer fractional states, including $\nu_{\text{bot}}+\frac{2(\nu_{\text{top}}-2)}{3}=1$, $\nu_{\text{bot}}+\frac{3(\nu_{\text{top}}-2)}{5}=1$, $\nu_{\text{bot}}+\frac{3(\nu_{\text{top}}-2)}{2}=1$, $\frac{3\nu_{\text{bot}}}{2}+(\nu_{\text{top}}-2)=1$.}
\end{figure}

\begin{figure}
\centering
\includegraphics{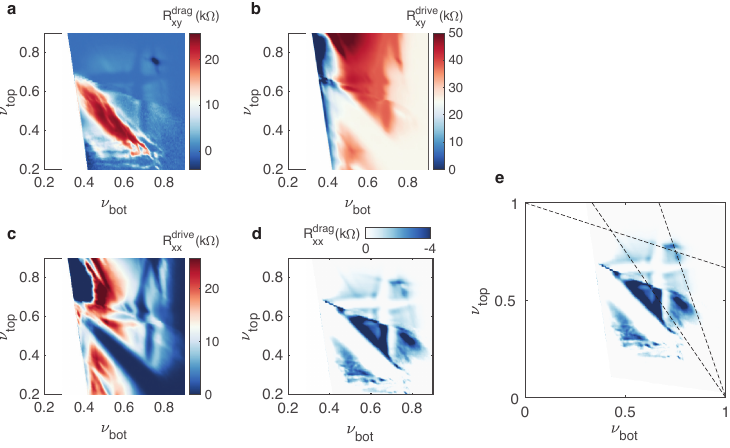}%
\caption{\label{fig:SI_16T_0505} Zoomed-in scans on the N = 0 interlayer states near \((\nu_{\text{bot}}, \nu_{\text{top}}) = (0.5, 0.5)\) at \(V_{\text{int}}=0\) and at 16~T. \textbf{a,} \(R_{\text{xy}}^{\text{drag}}\). \textbf{b,} \(R_{\text{xy}}^{\text{drive}}\). \textbf{c,} \(R_{\text{xx}}^{\text{drive}}\). \textbf{d,} \(R_{\text{xx}}^{\text{drag}}\), the same data as Fig.~2e in the main text. \textbf{e,} Dashed lines overlaid on top of the \(R_{\text{xx}}^{\text{drag}}\) data, marking interlayer fractional states, including $\nu_{\text{bot}}+\frac{2\nu_{\text{top}}}{3}=1$, $\nu_{\text{bot}}+\frac{\nu_{\text{top}}}{3}=1$ and $\frac{\nu_{\text{bot}}}{3}+\nu_{\text{top}}=1$.}
\end{figure}

\begin{figure}
\centering
\includegraphics{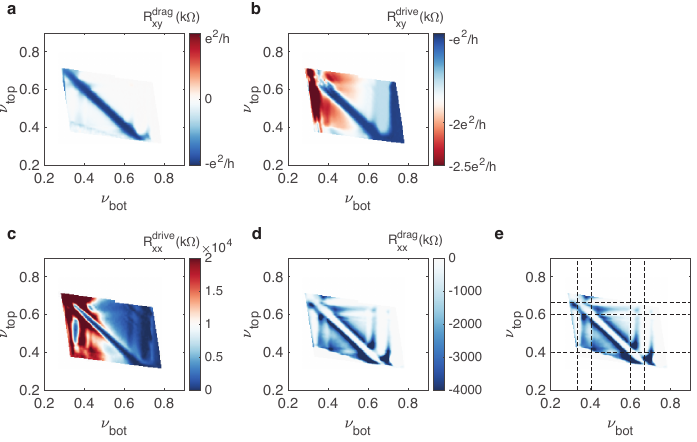}%
\caption{\label{fig:SI_25T_0505} Zoomed-in scans on the N = 0 interlayer states near \((\nu_{\text{bot}}, \nu_{\text{top}}) = (0.5, 0.5)\) at \(V_{\text{int}}=0\) and at 25~T. \textbf{a,} \(R_{\text{xy}}^{\text{drag}}\). \textbf{b,} \(R_{\text{xy}}^{\text{drive}}\). \textbf{c,} \(R_{\text{xx}}^{\text{drive}}\). \textbf{d,} \(R_{\text{xx}}^{\text{drag}}\). \textbf{e,} Dashed lines overlaid on top of the \(R_{\text{xx}}^{\text{drag}}\) data, marking the strong intralayer 1/3, 2/3, 2/5, 3/5 fractional quantum Hall states in both top and bottom BLG.}
\end{figure}

\begin{figure}
\centering
\includegraphics[width=0.8\textwidth]{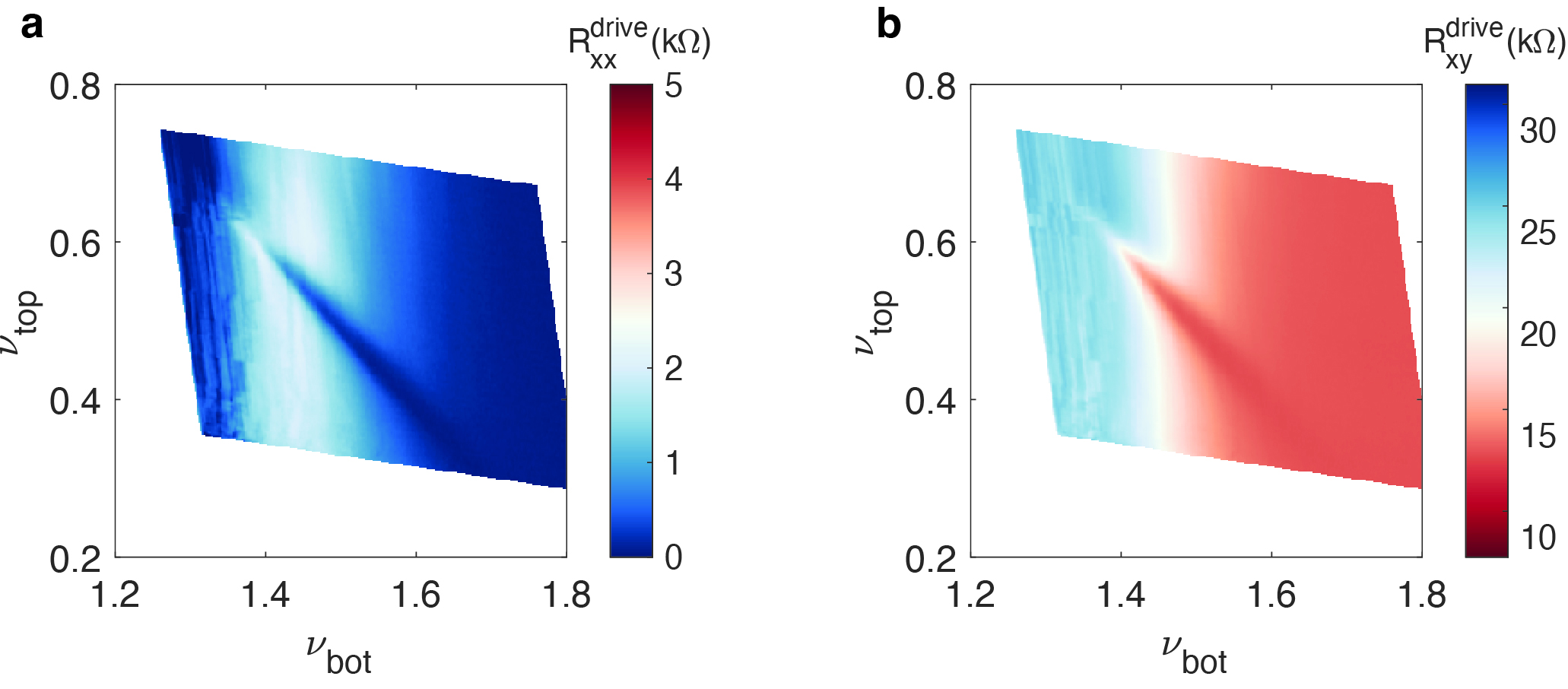}%
\caption{\label{fig:SI_double_bilayer_fig7} Zoomed-in scans on the N = 1 interlayer states near \((\nu_{\text{bot}}, \nu_{\text{top}}) = (1.5, 0.5)\) at \(V_{\text{int}}=0.1\)~V and at 16~T. \textbf{a,} \(R_{\text{xx}}^{\text{drive}}\), \textbf{b,} \(R_{\text{xy}}^{\text{drive}}\).}
\end{figure}

Fig.~\ref{fig:SI_16T_0525}–\ref{fig:SI_double_bilayer_fig7} are zoomed-in scans of various resistance channels for the $N = 0$ and $N = 1$ EC states, complementing the zoom-in scans shown in the main text Fig.~2 and 3. We observe fractional states of both intra- and interlayer nature. For example, the vertical streaks of $R_{\text{xx}}^{\text{drag}}=0$ at \(\nu_{\text{bot}}=\frac{2}{3}\) in Fig.~2c and 2e, as well as the horizontal streaks of $R_{\text{xx}}^{\text{drag}}=0$ at \(\nu_{\text{top}}=2+\frac{2}{3}\) in Fig.~2c and \(\nu_{\text{top}}=\frac{2}{3}\) in Fig.~2e correspond to intra-layer FQH states that developed independently in each BLG. There are also \(R_{\text{xx}}^{\text{drag}}\) peaks that follow fractional-slope lines corresponding to interlayer FQH states, including $\nu_{\text{bot}}+\frac{2(\nu_{\text{top}}-2)}{3}=1$ and $\nu_{\text{bot}}+\frac{3(\nu_{\text{top}}-2)}{5}=1$ in Fig.~2c, and $\nu_{\text{bot}}+\frac{2\nu_{\text{top}}}{3}=1$, $\nu_{\text{bot}}+\frac{\nu_{\text{top}}}{3}=1$ and $\frac{\nu_{\text{bot}}}{3}+\nu_{\text{top}}=1$ in Fig.~2e. At higher magnetic fields, additional intra- and interlayer FQH states emerge. For example, compared to 16~T (Fig.~2c), the data at 25~T (Fig.~2d) present additional intralayer FQH states at $\nu_{\text{top/bot}}=\frac{1}{3}, \frac{3}{5}$, and interlayer FQH states along $\frac{3\nu_{\text{bot}}}{2}+(\nu_{\text{top}}-2)=1$ and $\nu_{\text{bot}}+\frac{3(\nu_{\text{top}}-2)}{2}=1$. In particular, the dashed lines in all the e panels of Fig.~\ref{fig:SI_16T_0525}-\ref{fig:SI_16T_0505} illustrate the additional interlayer states discussed in the main text. These lines well define linear relations with fractional slopes, corresponding to interlayer fractional states or the so-called semiquantized states.

We note that near \((\nu_{\text{bot}},\nu_{\text{top}})=(0.5,0.5)\) at 16T (Fig.~2e in the main text and Fig.~\ref{fig:SI_16T_0505}) has an additional sharp transition in the middle of the structure. This occurs because the top BLG switches valley flavor here near zero $D_{\text{top}}$ while remaining in the $N = 0$ orbital. Although there is an overall drop of $R_{\text{xx}}^{\text{drag}}$, the general state structure remains unchanged.

\section{S6. Additional discussion on the interlayer bias dependence}

The idea behind using the \(V_{\text{int}}\) dependence to identify the existence conditions for the \(N=1\) interlayer EC states within the parameter space of \(\nu_{\text{top/bot}}\) and \(D_{\text{top/bot}}\) runs as follows. We first note that the \(V_{\text{int}}=0.07\)~V data shown in Fig.~3a and 3b, as well as the \(V_{\text{int}}=0.1\)~V data shown in Fig.~\ref{fig:SI_other_channel01}, confirms that these N = 1 EC states can occur at $\nu_{\text{bot}}=$ 3.5 or 1.5. Then we simultaneously vary all three gates \(V_{\text{TG}}\), \(V_{\text{BG}}\) and \(V_{\text{int}}\) while keeping the bottom filling fixed to $\nu_{\text{bot}}=$ 3.5 or 1.5. More precisely, we keep $V_{\text{BG}}+\frac{C_{\text{int}}}{C_{\text{b}}}V_{\text{int}}$ constant. The resulting 2D \(V_{\text{TG}}-V_{\text{int}}\) (or equivalently \(V_{\text{TG}}-V_{\text{BG}}\)) map varies \(\nu_{\text{top}}\), \(D_{\text{top}}\) and \(D_{\text{bot}}\), providing a direct visualization of how the $N = 1$ EC states form and evolve. It is more intuitive to convert the $V_{\text{TG}}-V_{\text{int}}$ map into an \(n_{\text{top}}-D_{\text{top}}\) map, effectively generating a phase diagram of top layer bilayer graphene, as shown in main text Fig.~4. 

\begin{figure}
\centering
\includegraphics{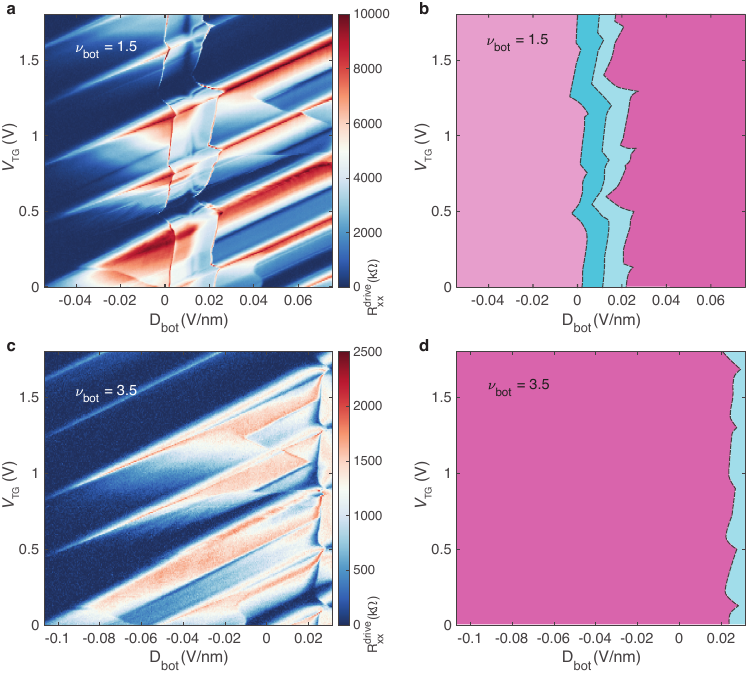}%
\caption{\label{fig:SI_bot_flavors} \textbf{a} and \textbf{c,} $R_{\text{xx}}^{\text{drive}}$ at fixed $\nu_{\text{bot}}=$ 1.5 and 3.5, respectively, as a function $D_{\text{bot}}$ and $V_{\text{TG}}$. These are the same data as Fig.~4c and 4a in the main text but plotted against different axes. \textbf{b} and \textbf{d}, Schematics of the flavor transition features in the bottom BLG. Color shows different orbital and valley flavors, with the same color coding scheme as Fig.4 in main text.}
\end{figure}

Here in Figs.~\ref{fig:SI_bot_flavors}a and c, we plot the same \(R_{\text{xx}}^{\text{drive}}\) data as Fig.~4a and c in the main text, but as a function of \(V_{\text{TG}}\) and \(D_{\text{bot}}\), so we can better identify the flavor transition features in the lower BLG. They can be easily identified as the features that are generally perpendicular to \(D_{\text{bot}}\) axis. Again, based on the phase diagram we obtained in Fig.~1e, here we similarly illustrate the bottom BLG LL in Figs.~\ref{fig:SI_bot_flavors}b and d, with each flavor colored following the same color scheme as in the main text, and dashed lines tracking the flavor transitions. Knowledge of the flavor distribution of the BLG confirms that the $N = 1$ EC states only appear when the N = 1 wavefunction is mostly populated in the top layer of the lower BLG.

\begin{figure}
\centering
\includegraphics{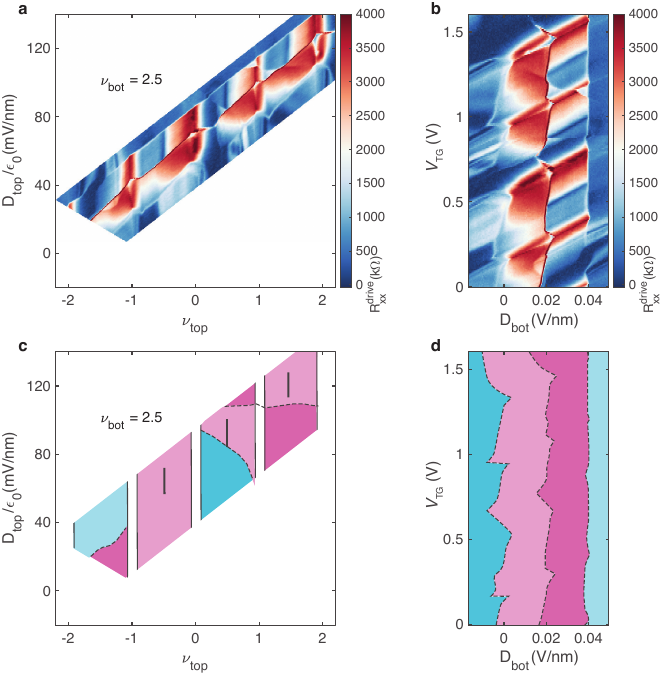}%
\caption{\label{fig:SI_flavors_line2} $R_{\text{xx}}^{\text{drive}}$ at fixed $\nu_{\text{bot}}=2.5$, \textbf{a,} as a function $D_{\text{top}}$ and $\nu_{\text{top}}$, and \textbf{b,} $D_{\text{bot}}$ and $V_{\text{TG}}$, respectively. \textbf{c}, Schematics of the features and flavors in the top BLG. \textbf{d}, Schematics of the features and flavors in the bottom BLG.}
\end{figure}

Fig.~\ref{fig:SI_flavors_line2} shows similar scans with the bottom bilayer graphene fixed to \(\nu_{\text{bot}} = 2.5\), again consistent with our observation of the flavor combination required for $N = 1$ EC. 

\section{S7. Disappearance of some N = 0 interlayer states at a finite interlayer bias}

\begin{figure}
\centering
\includegraphics[width = 0.4\textwidth]{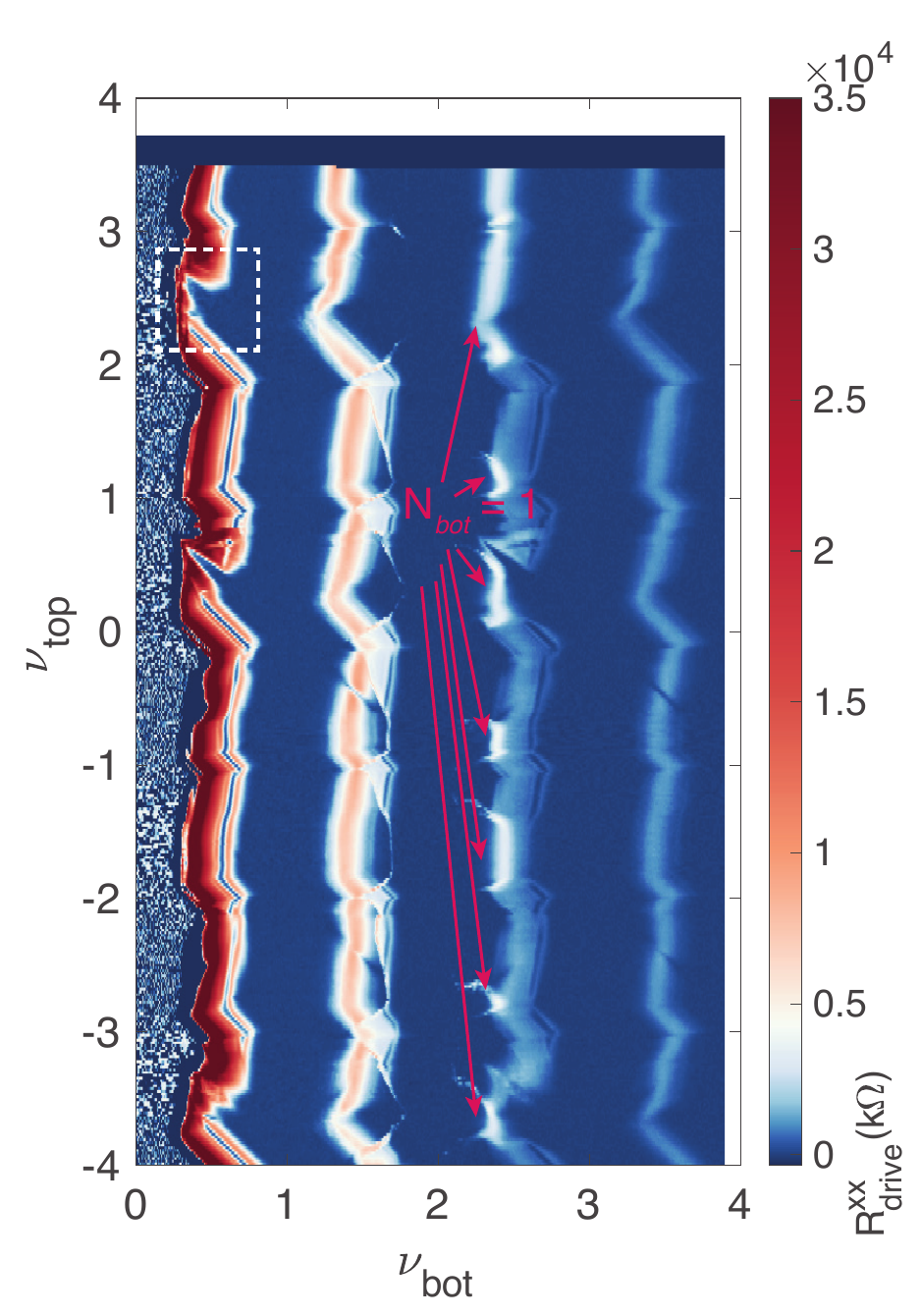}%
\caption{\label{fig:SI_transitions_within_N1} $R_{\text{xx}}^{\text{drive}}$ same as in main text Fig.~3a. Red arrows point to where N = 1 orbital appears in the mostly N = 0 orbital in \(2<\nu_{\text{bot}}<3\).}
\end{figure}

Compared to Fig.~2a,b at \(V_{\text{int}}=0\), the absence of N = 0 interlayer states at \((\nu_{\text{bot}}, \nu_{\text{top}})\sim\)  (0.5, -1.5) and (2.5, -1.5) in Fig. 3a,b at \(V_{\text{int}}=0.07\)~V is due to the mismatch of the orbital configuration, with one layer going into N = 1 orbital while the other remaining in N = 0. For example, at \(V_{\text{int}}=0.07\) V, when  top = -1.5,  \(D_{\text{top}}/\epsilon_0 \sim\)\ 6 mV/nm. Referring to the phase diagram in Fig. 1e, this corresponds to the top BLG being in N = 1 orbital. 
As for the other two locations where  \(\nu_{\text{top}} \) = 2.5,  \(D_{\text{top}}/\epsilon_0 \sim\) 120 mV/nm, which indeed corresponds to top BLG being in N = 0 orbital. However, at \((\nu_{\text{bot}}, \nu_{\text{top}})\)= (2.5, 2.5), the bottom layer enters the N = 1 orbital because near  \(\nu_{\text{bot}}\) =2.5,  \(D_{\text{top}}/\epsilon_0 \sim\)\ -17 mV/nm, which is close to the transition between N = 0 and N = 1. In fact, this transition is clearly visible in $R_{\text{xx}}^{\text{drive}}$ as shown in Fig.~\ref{fig:SI_transitions_within_N1}, in the range of \(2<\nu_{\text{bot}}<3\), whereas top changes, the bottom BLG orbital number oscillates between 1 and 0. This behavior appears as the alternating pattern of more resistive (white) and less resistive (dimmer) regions. For  \((\nu_{\text{bot}}, \nu_{\text{top}})\)= (0.5, 2.5), both layers are in the N = 0 configuration, and an interlayer state is therefore expected. Indeed the characteristic  $R_{\text{xx}}^{\text{drive}}$=0 diagonal feature is present (marked by the dashed square in Fig.~\ref{fig:SI_transitions_within_N1}). However, the corresponding drag signal is not observed at this region. While we cannot be fully certain, the proximity to the bottom BLG charge neutrality, where contact quality deteriorates, may play a role. Further systematic study with higher quality samples with improved contact quality may shed light on this observation.

\clearpage



%